\documentclass[10pt,conference,letterpaper]{IEEEtran}

\usepackage{amssymb}

\usepackage{balance}  
\usepackage{times,amsmath,graphics}
\usepackage{epsfig}
\usepackage{tabularx}
\usepackage{mathptmx,amsmath,epsfig,url}
\usepackage{times}
\usepackage{subfig}
\usepackage{subfloat}
\usepackage{multirow}
\usepackage[linesnumbered,ruled,algonl,vlined]{algorithm2e}

\newcommand{\eat}[1]{}
\newtheorem{definition}{Definition}[section]
\newcommand{\argmin}{\operatornamewithlimits{arg\,min}}

\newcommand{\scriptbelow}{\operatornamewithlimits{}}

\title{Scalable Visibility Color Map Construction in Spatial Databases}
\author{%
{Farhana Murtaza Choudhury{\small $~^{\#1}$}, Mohammed Eunus Ali{\small $~^{\#2}$}, Sarah Masud{\small $~^{\#3}$}, Suman Nath{\small $~^{*4}$, } Ishat E Rabban{\small $~^{\#5}$}}%
\vspace{1.6mm}\\
\fontsize{10}{10}\selectfont\itshape
$~^{\#}$Department of CSE, Bangladesh University of Engineering and Technology, Dhaka, Bangladesh\\
\fontsize{9}{9}\selectfont\ttfamily\upshape
$~^{1}$farhanamc@gmail.com,
$~^{2}$eunus@cse.buet.ac.bd,
$~^{3}$sarahmasud.sm@gmail.com,
$~^{5}$ieranik@yahoo.com
\vspace{1.2mm}\\
\fontsize{10}{10}\selectfont\rmfamily\itshape
$~^{*}$Microsoft Research, Redmond, Washington, USA\\
\fontsize{9}{9}\selectfont\ttfamily\upshape
$~^{4}$sumann@microsoft.com
}

\begin{document}
\maketitle
%
\maketitle

\begin{abstract}
Recent advances in 3D modeling provide us with real 3D
datasets to answer queries, such as \emph{``What is the best
position for a new billboard?"} and \emph{``Which hotel room has
the best view?"} in the presence of obstacles.
These applications require measuring and differentiating the visibility of an object (target) from different viewpoints in a dataspace, e.g., a billboard may be seen from two viewpoints but is
readable only from the viewpoint closer to the target. In this
paper, we formulate the above problem of quantifying the
visibility of (from) a target object from (of) the surrounding
area with a \emph{visibility color map} (VCM). A VCM is
essentially defined as a surface color map of the space, where
each viewpoint of the space is assigned a color value that denotes
the visibility measure of the target from that viewpoint.
Measuring the visibility of a target \emph{even} from a single
viewpoint is an expensive operation, as we need to consider
factors such as distance, angle, and obstacles between the
viewpoint and the target. Hence, a \emph{straightforward approach}
to construct the VCM that requires visibility computation for every
viewpoint of the surrounding space of the target, is
prohibitively expensive in terms of both I/Os and computation, especially for a real dataset comprising of thousands of obstacles. We propose an efficient approach to compute
the VCM based on a key property of the human vision that
eliminates the necessity of computing the visibility for a large
number of viewpoints of the space. To further reduce the
computational overhead, we propose two approximations; namely,
\emph{minimum bounding rectangle} and \emph{tangential} approaches
with guaranteed error bounds. Our extensive experiments demonstrate the
effectiveness and efficiency of our solutions to construct the VCM
for real 2D and 3D datasets.

\eat{Our approach results in \emph{three orders} and \emph{five
orders} of magnitude less computation than that of the
straightforward approach for real 2D and 3D datasets,
respectively. To further reduce the computational overhead, we
propose two approximations namely \emph{minimum bounding
rectangle} and \emph{tangential} approaches with guaranteed error
bounds. Our experiments demonstrate that on average \emph{minimum
bounding rectangle} and \emph{tangential} approaches are 50\% and
15\% faster while introducing 8\% and 5\% error, respectively, in
comparison with our exact method.}

\eat{Our approach results in at least \emph{three orders} of
magnitude less computation than that of the straightforward
approach. To further reduce the computational overhead, we propose
two approximations namely \emph{minimum bounding rectangle} and
\emph{tangential} approaches with guaranteed error bounds. Our
experiments demonstrate the effectiveness and efficiency of our
solutions to construct the VCM for real 2D and 3D datasets.}
\end{abstract}

\section{Introduction}
Recent advances in large-scale 3D modeling have enabled capturing urban environments into 3D models. These 3D models give photo-realistic resembling of urban
objects such as buildings, trees, terrains etc. and are widely used by popular 3D mapping services, e.g., Google Maps, Google Earth, and Bing Maps.
The increasing availability of these realistic 3D datasets provides us an opportunity to answer many real-life queries involving visibility in the presence of 3D obstacles. For example, a tourist may check the visibility of a new year firework event from different locations in the surrounding areas so that he can pick a good spot to enjoy it; an apartment buyer may want to check visibility of near-by sea-beach and mountains from various available apartments; and an advertising company may wish to determine visibility of their existing billboards from surrounding areas and find a suitable location for a new billboard accordingly.

In this paper, we investigate efficient techniques to answer the underlying query required by the above applications: computing visibility of an object (e.g., firework event, billboard) from the surrounding continuous space, or that of the surrounding space from a source viewpoint. Our target applications treat visibility as a continuous notion---e.g., a billboard may be more visible from one location than another, depending on factors such as distance, viewing angle, and obstacles between the viewpoint and the target. We therefore use a {\em visibility function} that provides real-valued visibility measures of various points in the (discretized) 3D space, where the visibility measure of a point denotes its visibility from the viewpoint or to the target object. Thus, the answer to our target query is essentially the visibility measures for every point in the 3D space. The result can be graphically represented as a {\em heat map}, by assigning colors to various points according to their visibility measures. We call this a {\em visibility color map} (VCM) of the space for a given target or for a given viewpoint.

Recent works have shown how database techniques can enable efficiently answering various types of visibility queries in the presence of obstacles. Various nearest neighbor (NN) queries consider visibility of objects~\cite{SaranaVNN,Gao:2009:CON:1559845.1559906,Gao:2009:CVN:1516360.1516378}; for example, the visible nearest neighbor query~\cite{SaranaVNN} finds the nearest neighbors that are visible from the source. However, these works, like various other computer graphics works~\cite{vmapDefinition,bittnerOccusntree,grasset1999accurate,keeler2007spherical,bittnerWonka2003visibility,stewart1998computing}, treat visibility as a binary notion: a point is either visible or not from another point. In contrast, in our target applications, visibility is a continuous notion. \eat{Recently, Masud et al. proposes techniques for computing continuous visibility measure between two given points in the 3D space (e.g., computing visibility of a billboard from a given location)~\cite{MOV}. In contrast to such {\em point-to-point} visibility solutions, our target applications require visibility to or from a continuous space.}Recently, Masud et al. proposed techniques for computing continuous visibility measure of a target object from a particular point in 3D space (e.g., computing visibility of a billboard from a given location)~\cite{MOV}. On the contrary, our target applications require visibility calculation from or of a continuous space where there is no specific viewpoint.

One straightforward way to generate a VCM is to descretize the 3D
space and to use the techniques in~\cite{MOV} to compute
visibility measure for each discrete point in the space. However,
this can be prohibitively expensive. For example, discretizing the
surrounding space into 1000 points in each dimension would give a
total of $10^9$ points in the 3D space; and computing visibility
measure for each point by using techniques in~\cite{MOV} would
take 128 days! The huge cost comes from two sources: (i)
computing the visibility measure based on the distance and angle
from all viewpoints, which is computationally expensive and (ii)
accessing a large set of obstacles from the database, which is I/O
expensive.

We address the above challenges with a three-step solution that
uses several novel optimizations to reduce computational and I/O
overhead. First, we partition the dataspace into a set of {\em
equi-visible cells}, i.e., all points inside a cell have equal
visibility of the target object in terms of visual appearance. We
exploit the key observation that when a lens (e.g., a human eye)
sees an object without any obstacles, it cannot differentiate
between its visual appearances from a spatially close set of
points within an angular resolution (or spatial resolution) of
$\approx 4$ arcminutes ($\approx 0.07$
degrees)~\cite{angResolutionHuman}. Thus, we can safely prune the
visibility computation for a large number of viewpoints within the
angular or spatial resolution without affecting viewer's
perception. This optimization significantly reduces the
computation cost, as we can compute only one visibility measure
for each cell.

In the next step, we consider the effect of obstacles. We compute
{\em visible regions}, the regions in the space from where the
target object is completely visible in the presence of obstacles. In the
final step, we assign visibility measures to these regions from the
corresponding cells by spatial joins. Both steps are I/O and computation
intensive. For example, they both require retrieving a large
number of cells and obstacles from the spatial database. To reduce
I/O costs, we employ various indexing techniques to incrementally
retrieve a small number of obstacles and cells near the target
object. These steps also require performing many computationally
expensive operations such as polygon intersections of
irregular-shaped regions and cells. To reduce such costs, we
represent regions with regular shaped polygons in a quad-tree. We
also propose two approximations that further reduce the cost while
providing guaranteed small error bounds.



We have evaluated the performance of our solution with real 3D
maps of two big cities. We compare our solution with a \emph{baseline} approach that divides the space into a regular shape grid of 500 cells in each dimension and computes visibility from each grid cell. The baseline approach results into more than 30\% error while requiring about 800 times more computation time and six orders of magnitude more I/O than our solution. Hence, in the baseline approach, dividing the space into more cells for more accuracy is not feasible for practical applications. On the other hand, our approach provides efficient and effective solution.

\eat{The results show that, compared to the
baseline solution described before, our solution reduces the
computation by four- to five-orders of magnitude for both
datasets. The two approximation algorithms further speed up the
process by 54\% and 23\% while introducing 9\% and 5\% average
errors, respectively.}

In summary, we make the following contributions:
\begin{itemize}
    \item We formulate the problem of efficiently constructing a visibility color map (VCM) in the presence of obstacles in 2D and 3D spaces.
    \item We propose an efficient solution to construct a VCM. The solution uses various novel optimizations to significantly reduce the computational and I/O overhead compared to a baseline solution.
    \item We propose two approximations with guaranteed error bounds and reduced computation to construct the VCM.
    \item We conduct extensive experiments in two real datasets to show the effectiveness and efficiency of
    our approaches.
\end{itemize}

\section{Problem Formulation and Preliminaries}

\subsection{Problem Formulation}
\label{sec:pf} The construction of a \emph{visibility color map}
(VCM) can be seen from two perspectives: \emph{target-centric VCM}
and \emph{viewer-centric VCM}. The construction of a
\emph{target-centric VCM} involves determining how much a given
target is visible from every point\footnote{We have used
point and viewpoint interchangeably.} of the space. On the other
hand, a \emph{viewer-centric VCM} involves determining how much visible each point in the surrounding space is from a given viewer's location.

In both cases, we need to compute visibility and
produce a color map of the space where each point of the space is
assigned a color value that corresponds to the visibility measure of that point.

\begin{definition}
\emph{Visibility Color Map (VCM)}

Given a $d$-dimensional dataspace $\mathbb{\Re}^d$ and a set $O$
of obstacles, the VCM is a color map $C$, where for each point
$\textsc{x} \in \mathbb{\Re}^d$, there exists a color
$c_{\textsc{x}} \in [0,1]$. The color $c_{\textsc{x}}$ corresponds
to the visibility of a given target object $T$ \emph{from}
$\textsc{x}$ in case of \emph{target-centric VCM} and the
visibility \emph{of} $\textsc{x}$ from a given viewer's location $q$
in case of \emph{viewer-centric VCM}. Here, $c_{\textsc{x}}$ is
normalized between $0$ and $1$.
\end{definition}

Without loss of generality, we limit our discussion to the
construction of target-centric VCM in the subsequent
sections. However, our solution is also applicable to the
viewer-centric VCM construction, as explained in
Section~\ref{sec:viewer}.

\eat{So we specify a metric for visibility measurement in the
following section.}

The core of computing a VCM is computing visibility of the target
from various points in the space. The most common measure of
visibility (or, the perceived size) of an object is the visual
angle~\cite{baird1970psychophysical}, which is the angle imposed
by the viewed object on a lens. The visual angle mainly depends on
the characteristics of the viewing lens as well as the distance,
angle, and obstacles between the viewer and the
target~\cite{visibilityFactors}.


\subsection{Factors Affecting Visibility}
\label{subsec:vp}

\begin{figure}
\label{fig:Parameters}
\centering
\includegraphics[height=.9in]{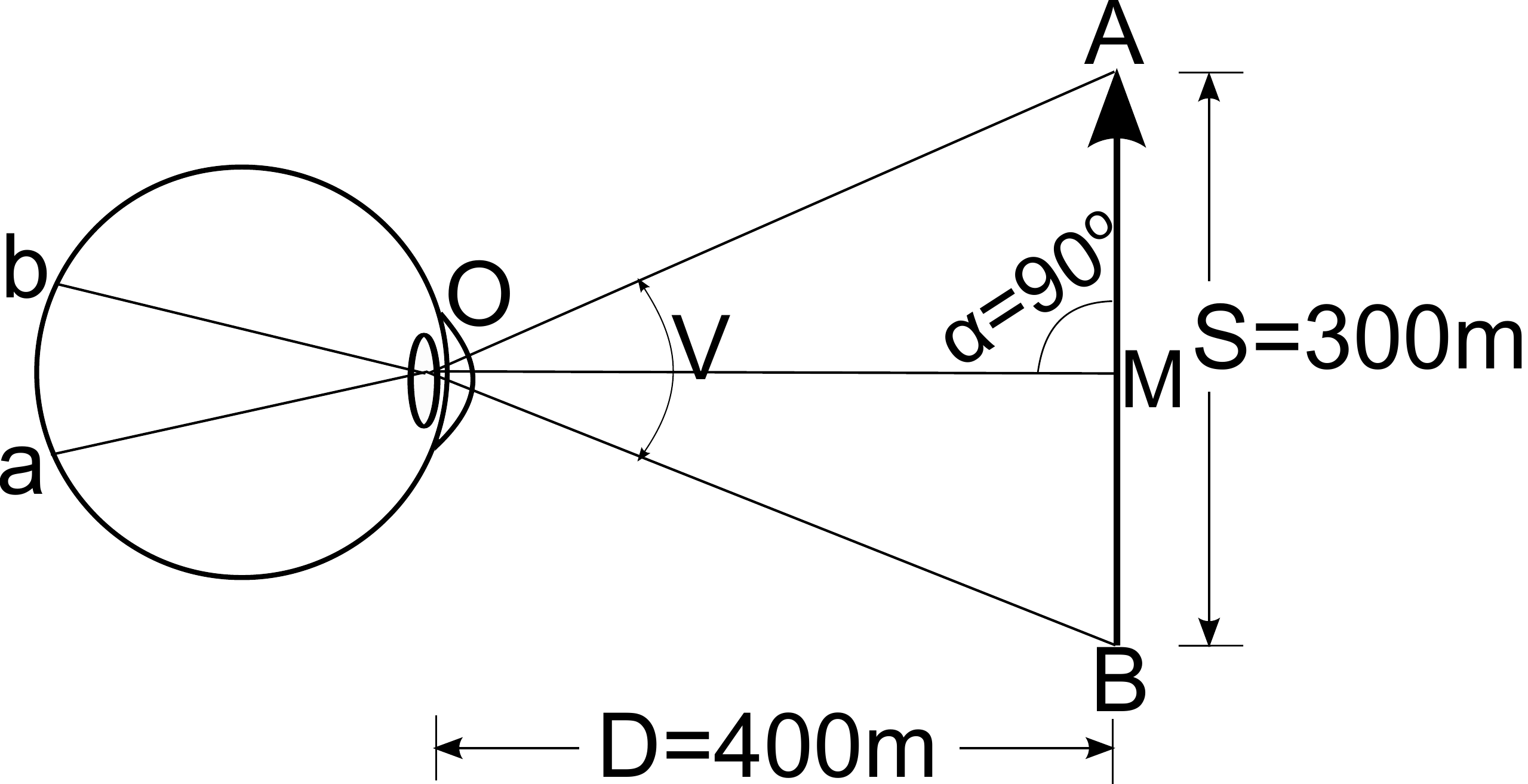}
\caption{Effect of distance and angle on visibility}
\label{fig:Parameters}
\vspace{-7.6mm}
\end{figure}

\subsubsection{Relative Position of the Lens and the Target}
\label{subsubsec:pos} The perceived visibility of a target object
mainly depends on the relative position the viewing lens and the
target. For a specific target, the visibility varies with the
change of distance and angle between the lens and the target.

\textbf{\emph{Distance:}} If the distance between a target $T$ and
a lens increases, the perceived size of $T$ becomes smaller. This
is because the visual angle imposed by an object decreases with
the increase of the distance between that object and the viewer.
As shown in Fig.~\ref{fig:Parameters}, $AB$ is a target
object of length $S$ and the position of a lens is $O$. When the
midpoint of $AB$ is at an orthogonal distance $D$ from $O$, the
visual angle $V$ is calculated using the following formula~\cite{kaiser1996joy},

\vspace{-2.8mm}
\begin{equation}
\label{eq}
V = 2arctan(\frac{S}{2D})
\end{equation}
\vspace{-4.8mm}

\textbf{\emph{Angle:}} The perceived size of a target $T$ depends
on the angle $\alpha$ between the lens and $T$~\cite{engineeringDrawinig}. If an object is viewed from an
oblique angle, the perceived size of that object becomes smaller
than the original size. For the equidistant positions of the lens
from $T$, the visual angle $V$ varies for different values of
$\alpha$. Let $AB$ be a line of length $S$. A line $\overline{qm}$
connecting a point $q$ and the midpoint $m$ of $AB$, creates an
angle $\alpha$ with $AB$. If $\alpha = 90^o$, the perceived size
of $AB$ from a nominal distance is same as the original length,
$S$. Otherwise, according to the concept of oblique projection~\cite{engineeringDrawinig}, the
perceived length $S_\alpha$ of $AB$ from $q$ is

\vspace{-5.4mm}
\begin{equation}
\label{eq2} S_\alpha = \frac{\displaystyle \alpha}{\displaystyle
90^o} \times S
\end{equation}
\vspace{-5mm}

\eat{This is depicted in Figure~\ref{fig:Parameters}(b), where a target
object $AB$ of length $300m$ is viewed from a point $O$ located at a
distance of $400m$. The midpoint of $AB$ is $M$. As the line
$\overline{OM}$ creates $90^o$ angle ($\alpha$) with $AB$, the
perceived length of $AB$ from $O$ is $300m$. If $\alpha= 45^o$,
then the perceived length of $AB$ from $O$ would be $300m \times
\frac{\displaystyle 45^o}{\displaystyle 90^o} = 150m$.}

Thus, if we consider the effects of both distance and angle between the target and the lens, the visual angle $V$ can be expressed as $V = 2arctan(\frac{S_{\alpha}}{2D})$. In
Fig.~\ref{fig:Parameters}, the visibility of the target $AB$ of length $S$
is the visual angle $V$ imposed at the lens $O$. As in this case
$\alpha= 90^o$, so the visual angle $V$ is
$2arctan(\frac{\displaystyle 300m}{\displaystyle 2 \times 400m}) =
41.11^o$. If $\alpha= 45^o$, then $V$ is
$2arctan(\frac{\displaystyle 150m}{\displaystyle 2 \times 400m}) =
21.23^o$.

Besides the relative position of the lens, the visibility of a target is also affected by the presence of obstacles.


\subsubsection{Obstacles}
\label{subsec:vo}

\eat{A point $p$ on the surface of the target $T$ is invisible from a viewpoint $q$ in the dataspace if a
an obstacle obstructs the line of sight from $q$ to $p$. Thus the target $T$ is completely visible from $q$, if

If for a We term the set of viewpoints in the dataspace from where $T$ is not fully
visible as the \emph{obstructed region}.

and color them with the
value allocated for zero visibility (i.e., invisible). The
obstructed region is defined in terms of point to point
visibility.
}

To show the effect of obstacles on the visibility, we first define the term point to point
visibility.

\begin{definition}
\emph{Point to point visibility.}
Given two
points $p,p^\prime$ and a set $O$ of obstacles in a space, $p$ and $p^\prime$
are visible to each other if and only if the straight line connecting them,
$\overline{pp^\prime}$, does not cut through any obstacle, i.e., $\forall o \in
O,$ $\overline{pp^\prime} \cap o = \oslash $.
\end{definition}

Based on the definition of point to point visibility, we formally
define the obstructed region as:

\begin{definition}
\emph{Obstructed region.} Given a set $O$ of obstacles, a bounded
region $R$, and a target $T$, the obstructed region is the set of
points where for each point $p$, (i) $p$ is in $R$ and (ii) $p$ is not point
to point visible to \emph{all} points of $T$.
\end{definition}

The obstructed region contains viewpoints from where the target
object is not completely visible. Thus, we only need to measure the
visibility for the viewpoints residing outside the obstructed region that form the \emph{visible region}, and assign colors
to these viewpoints of the visible region according to the defined \emph{visibility measure} (i.e., visual angle).


\section{Constructing a VCM}
\label{sec:approach}
To construct a VCM for a given target $T$, we need to determine the visibility of $T$ from all discrete points in the surrounding space $\mathbb{\Re}$ in the presence of a set $O$ of obstacles. We represent visibility of each point $\textsc{x} \in \mathbb{\Re}$ with a color $c_{\textsc{x}}$, which is proportional to the visibility of $T$ from $\textsc{x}$. We use the terms {\em visibility measure} and {\em color} interchangeably.

One na\"{\i}ve approach to compute a VCM is to compute visibility
of $T$ from every single viewpoint $\textsc{x} \in \mathbb{\Re}$.
Depending on how finely we discretize the space $\mathbb{\Re}$,
there can be a large number of points, making the process
prohibitively expensive. \eat{For example, in a 3D space, having mere
1000 points in each dimension results in a total of $10^9$ points.
As our experiments show, computing visibility of these many points
takes 3083 hours! This is not acceptable in many real-world
scenarios.}The high overhead of this na\"{\i}ve approach comes due to the
expensive visibility computation from a large number of points and
expensive I/O operations to retrieve a large collection of
obstacles from a spatial database.

To address these problems, we propose an
efficient solution to construct a VCM. The key insights of our
solution come from the following two observations. First, human
eye cannot visually differentiate a target from viewpoints
in close proximity of each other, which eliminates the necessity
of computing visibility for all viewpoints in the surrounding space.
Second, in most cases, only a small subset of obstacles affect
visibility of the target, and thus retrieval of all obstacles is redundant. Such
redundancy can be avoided by using various indexing techniques.


In the rest of the section, we describe how we exploit these
observations and propose an efficient approach to compute a VCM
with reduced computational and I/O overhead. Our approach consists
of three steps:
\begin{enumerate}
    \item First, we partition the space into several {\em equi-visible cells} in the absence of obstacles exploiting the limitations of human vision. This enables us to compute one single visibility measure for each cell. This significantly reduces computational overhead in contrast to computing the color of each discrete viewpoint (Section~\ref{sec:parition}).

    \item Second, we compute the effect of obstacles and divide the surrounding space into a set of {\em visible regions} such that the target is completely visible from a viewpoint if and only if it is within a visible region. To reduce the I/O overhead, we index all obstacles and incrementally retrieve only the potential obstacles that can affect the visibility of the target (Section~\ref{sec:obstacles}).

    \item Finally, we join colors computed in the first step (that ignores obstacles) and visible regions computed in the second step to compute a VCM, i.e., colors of different parts of the space in the presence of obstacles. To reduce I/O overhead of retrieving results from the first two steps, we employ various indexes (Section~\ref{sec:merging}).
\end{enumerate}

For ease of explanation, we assume a 2D space and a target with
the shape of line in the subsequent sections. However, our
approach is applicable to any target shapes in 2D and 3D spaces.
Table~\ref{table:notations} lists the notations that we use.

\begin{table}
\centering \caption{Notations used and their meanings}
\label{table:notations}

\begin{small}
\begin{tabular}{|l|p{6cm}|}
\hline
{\bfseries Notation} & {\bfseries Meaning} \\
\hline
$T$ & The target object.\\
\hline
$O$ & A set of b obstacles $O$=$\{o_1,o_2,\dots,o_b\}$.\\
\hline
$V$ & The visual angle imposed in a lens by $T$.\\
\hline
$\mu$ & The angular resolution of a lens.\\
\hline
$\alpha$ & The angle between $T$ and the line connecting the lens and the midpoint of $T$.\\
\hline

\end{tabular}
\end{small}
\vspace{-15.5pt}
\end{table}

\subsection{Partitioning Space into Equi-visible Cells}
\label{sec:parition}
As mentioned before, the ability of a human eye (or a lens) to
distinguish the variation of small details of a target $T$ of size
$S$ is limited by its angular resolution $\mu$. To exploit this observation,
we partition the space into a set $\zeta$ of $n$ {\em equi-visible cells}
\{$\zeta_1,\zeta_2,\dots,\zeta_n$\}. Each cell $\zeta_i$ is constructed in a way so that
the deviation in visibility of $T$ from the viewpoints inside a
cell, measured as visual angle $V$, is not visually perceivable.
Hence for any two points $p,p^\prime \in \zeta_i$, visibility from $p$ and $p^\prime$ is perceived as same if,

\vspace{-5mm}
\begin{equation}
|V_p - V_{p^\prime}|\scriptbelow_{p,p^\prime \in \zeta_i} \le \mu
\vspace{-3mm}
 \end{equation}

Note that visibility is a {\em symmetric} measure: visibility of a target at location $p$ from a viewer (i.e., the lens) at
location $q$ is the same as that of a target at $q$ from location $p$. Thus, visibility of a target $T$
from the surrounding space is the same as visibility of the space from $T$'s location. Therefore, in computing visibility of $T$ from the space, we consider the viewer at the target's location and compute visibility of the space from that location.

Since the value of visual angle depends on the distance and angle
between the lens and the target, the partitioning is done in two
steps: distance based partition and angle based partition.

\subsubsection{Distance Based Partition}
\label{subsubsec:parition}
As the perceived size of $T$ varies with the change of distance
between $T$ and the viewer's location, our aim is to find a set
$D$ of $m$ distances, $D=\{d_0,d_1,\dots,d_{m-1}\}$, where for
each pair of points $p,p^\prime$ between $d_i$ and $d_{i+1}$, $0
\le i < m-2$, the variation of the perceived visibility from $p$
and $p^\prime$ is less than or equal to the angular resolution
$\mu$. Note that, since visibility varies as $\alpha$ deviates
from $90^o$ (as explained in Section~\ref{subsubsec:pos}), we set
$\alpha = 90^o$ as the default value for the distance based
partitioning.

\begin{figure}
\centering
\includegraphics[width=2.8in]{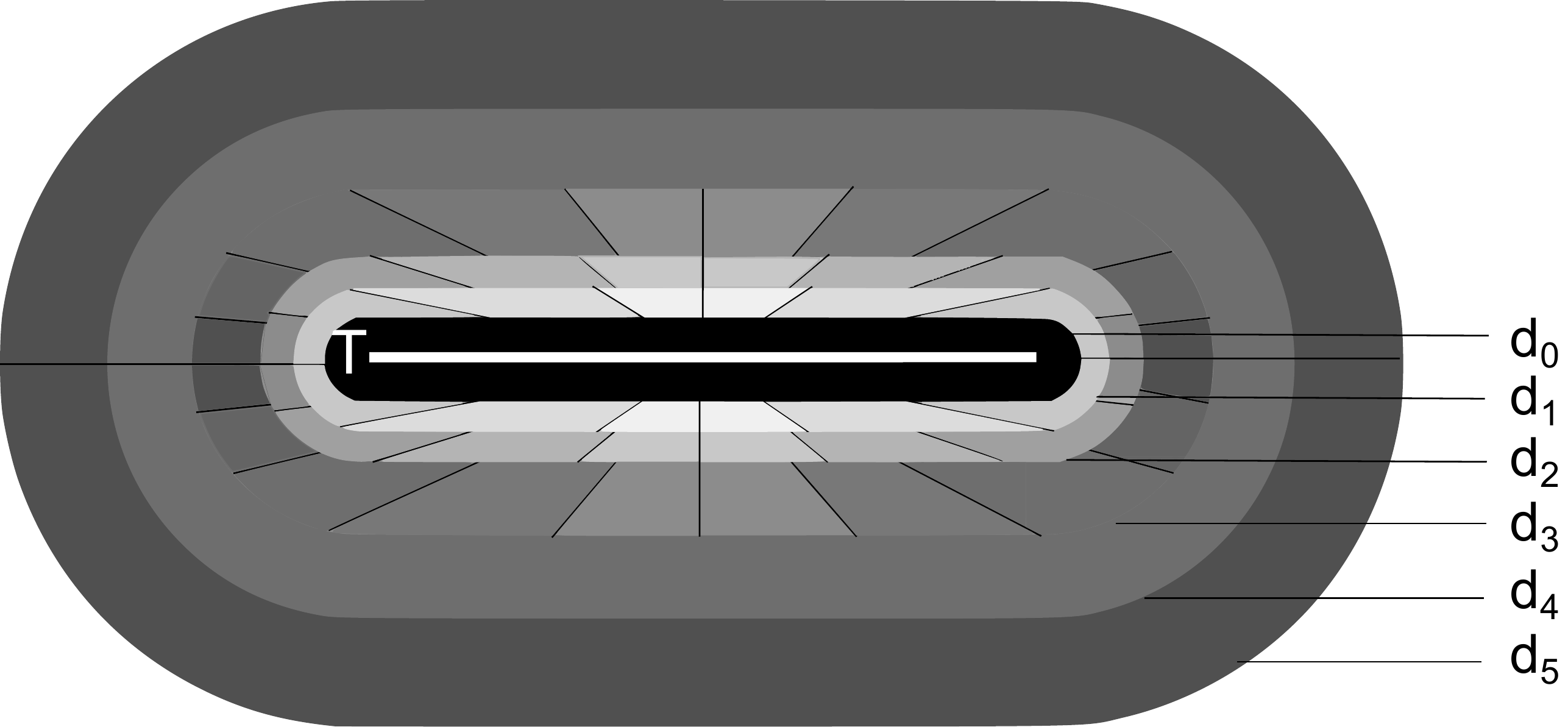}
\caption{Partitioning based on distance and angle}
\label{fig:partition}
\vspace{-22pt}
\end{figure}

Partitioning starts from the near point distance $d_0$, as a lens
cannot focus on any object that is nearer than $d_0$~\cite{nearPoint}. Initially,
the visual angle $V_0$ from the distance $d_0$ is calculated using
Equation~\ref{eq}. Then, starting from $V_0$, the value of the
visual angle is decreased by the amount of $\mu$ at each step and
the corresponding $d_i$ is calculated. When the imposed visual
angle from a distance $d_{m-1}$ is less than $\mu$, the distance
based partitioning process terminates as for any point farther
than $d_{m-1}$ (i.e., $d_{max}$), the perceived visibility is indistinguishable to
the viewing lens. So, we have a set $\{d_0,d_1,\dots,d_{m-1}\}$ of
distances where every range $<d_i,d_{i+1}>$ is a distance based
partition. Fig.~\ref{fig:partition} shows the
distance based partitions for a target $T$, where $d_0$ is the
near point distance and the distance based partitions are $<d_0,d_1>$,
$<d_1,d_2>$, and so on.

At this stage, we assign a single color for every distance based partition $<d_i,d_{i+1}>$. However, every point in a distance based partition does not perceive the same visibility of the target, e.g., two viewpoints at a same distance partition may have different perceived visibility due to different viewing angles. Thus, in the next section we incorporate the effect of viewing angle and partition the space into equi-visible cells.

\eat{Based on our discussion made so far, it may appear that two observers located at $q_1$ and $q_2$ have the same perceived visibility of the target as they fall in same distance based partition (Fig.~\ref{fig:dist}). But this is unrealistic as we ignore the effect of angle created between the target and the viewpoint. So in the next section we describe how the effect of angle is incorporated to partition the space into equi-visible cells.}

\subsubsection{Angle Based Partition}
\eat{Visibility of $T$ also depends on the angle $\alpha$ between
the lens and the target. For this reason, each partition formed by
the distance based partition is further partitioned according to
the variation in $\alpha$. }
For each distance based partition
$<d_i,d_{i+1}>$, $0 \le i < m-2$, we get $l_i$ numbers of angle
based partitions $<\gamma_{i,j}, \gamma_{i,j+1}>$, $0 \le j < l_i
-2$, where the value of $l_i$ is different for each distance based
partition. The visibility of every point of a partition
$<d_i,d_{i+1}, \gamma_{i,j}, \gamma_{i,j+1}>$ is considered as the
same. We call such a partition an \emph{equi-visible cell} (or, just {\em cell} in short).

\eat{Let $q$ be the location of a lens and $AB$ be a target object
of length $S$. A line $qm$, connecting the midpoint of $AB$ with
$q$ creates an angle $\alpha$-degree with $AB$. Given the distance
between target and viewer remains unchanged, the perceived length
of an object is same as the original length when $\alpha = 90^o$.}
As the change in perceived length due to change in $\alpha$ is
symmetric with respect to both the parallel and normal axes of the
object plane, we compute the angle based partitions only for the
first quadrant. The partitions in other quadrants are then
obtained by reflecting the partitions of the first quadrant. The
procedure of angle based partitioning is as follows:

(i) For each distance based partition $<d_i,d_{i+1}>$, the angle
based partitioning starts by initializing $\alpha_0 = 90^o$,
$\gamma_{i,0}=90^o$, and $S_0 =S$, where the size of $T$ is $S$.
The visual angle $V_0$ is calculated for $\alpha_0$ using
Equation~\ref{eq} where $D= d_i^\prime$. Here $d_i^\prime$ is the
average value of distances $d_i$ and $d_{i+1}$, i.e.,
$d_i^\prime = (d_i + d_{i+1})/2$.

(ii) At each step $j$ of the angle based partitioning for
$<d_i,d_{i+1}>$, the perceived size $S_j$ is calculated for visual
angle $V_j = V_{j-1} - \mu$ and distance $d_i^\prime$ using
Equation~\ref{eq}, where $j \ge 1$, e.g., $S_1 = 2D \times
tan\frac{V_1}{2}$, for $j=1$. The angle $\alpha_j$, for which
$S_j$ is perceived, is computed using Equation~\ref{eq2}, e.g.,
$\alpha_1 = \frac{\displaystyle S_1 \times 90^o}{\displaystyle S_0}$, for $j=1$. The visual
angle $V_j$ is obtained for the change in $\gamma_{i,j-1}$ by the
amount of $\alpha_j$, so $\gamma_{i,j}$ is updated as
$\gamma_{i,j-1} -\alpha_j$. Thus we get an angle based partition
$<\gamma_{i,j-1}, \gamma_{i,j}>$ for $<d_i,d_{i+1}>$ at each step.

(iii) When the perceived visual angle $V_j < \mu$, the angle based
partitioning process for a distance based partition terminates.

The above process is repeated for each distance based partition
and finally we get the set
$\zeta=\{\zeta_1,\zeta_2,\dots,\zeta_n\}$ of $n$ cells where $n =
m \times l_i$, $0 \le i < m -1$. Fig.~\ref{fig:partition} shows
the angle-based partitions for three distance-based partitions
$<d_0,d_1>$, $<d_1,d_2>$, and $<d_2,d_3>$ for a target object $T$.

After both partitioning steps are done, we compute visibility (i.e., color) of each cell. Since all viewpoints within a cell have the same visibility, we assign the color of the entire cell as the visibility of the center of the cell to the target $T$. \eat{We use the visibility metric described in Section~\ref{subsec:vm}.}


Note that we have not considered the effects of obstacles yet. A caveat of this is that in the presence of obstacles, the target $T$ may not be visible from an entire cell or parts of a cell even if the cell is assigned a good visibility value. We address this next by considering the effect of obstacles.


\subsection{Computing the Effects of Obstacles}\label{sec:obstacles}
Given a target object $T$ and a set $O$ of obstacles, we would like to determine the set $S$ of {\em visible regions} surrounding $T$ such that $T$ is completely visible from a viewpoint $q$ if and only if $q$ is inside a visible region.\eat{ Later we combine this information with the cell visibility information computed in the previous section.}

A na\"{\i}ve approach to determine the effects of obstacles is to retrieve all obstacles of $O$ and calculate the corresponding changes in the visibility. But this approach is prohibitively costly in terms of both I/O and computation, especially in the presence of a large number of obstacles. Moreover, considering all obstacles in the database is wasteful as only a relatively small number of obstacles around $T$ affect visibility. To efficiently retrieve this small number of obstacles around $T$, we index all obstacles in an R*-tree~\cite{rstar_tree}, a variation of R-tree~\cite{r_tree}. An R*-tree consists of a hierarchy of \emph{Minimum Bounding Rectangles (MBRs)}, where each MBR corresponds to a tree node and bounds all the MBRs in its sub-tree. Data objects (obstacles, in our case) are stored in leaf nodes.

%

\begin{figure}
\label{fig:obs_region}
\centering \subfloat[]{\includegraphics[width=1.65in]{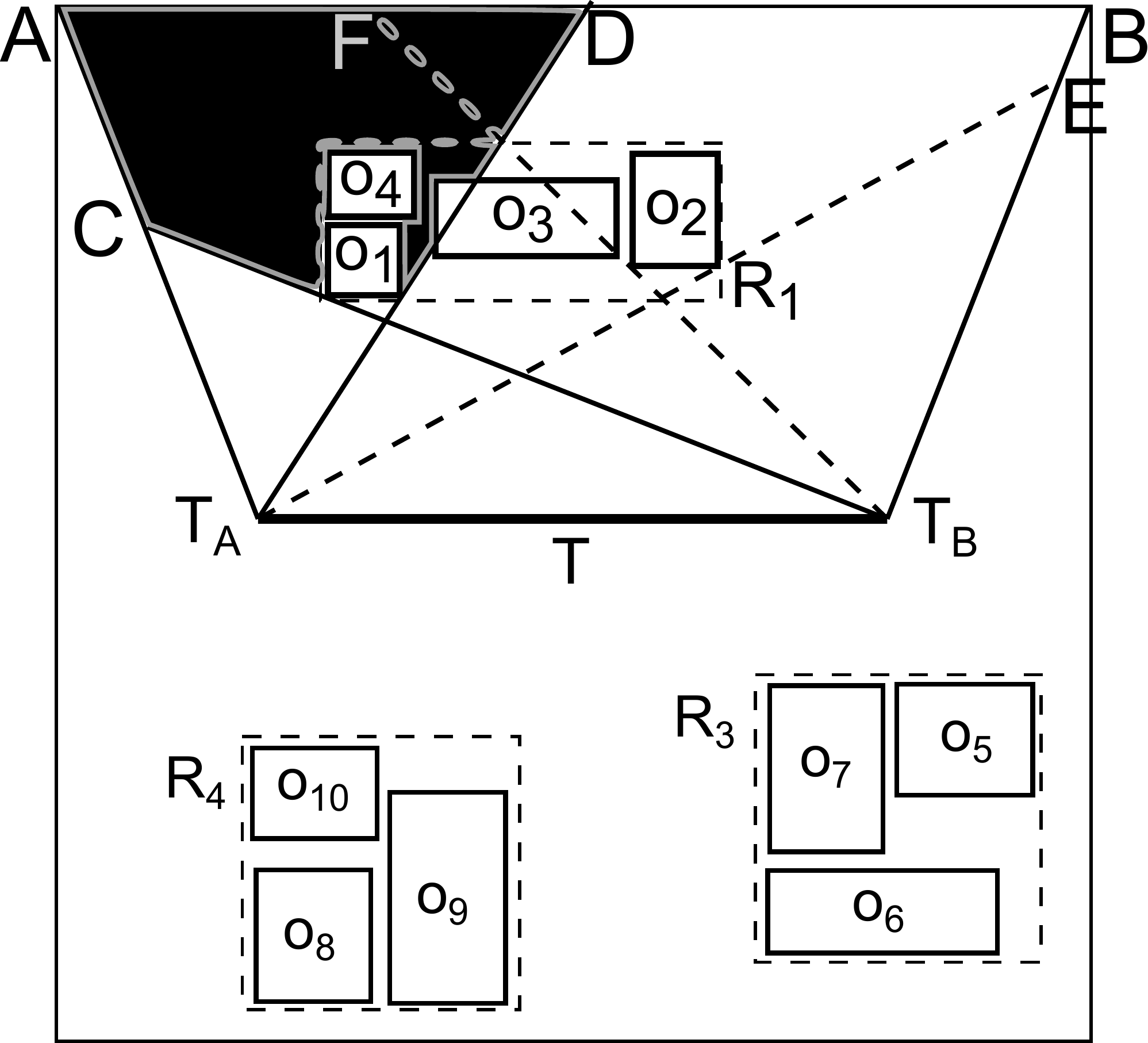}\label{fig:obs_region1}}
\hfill \subfloat[]{\includegraphics[width=1.65in]{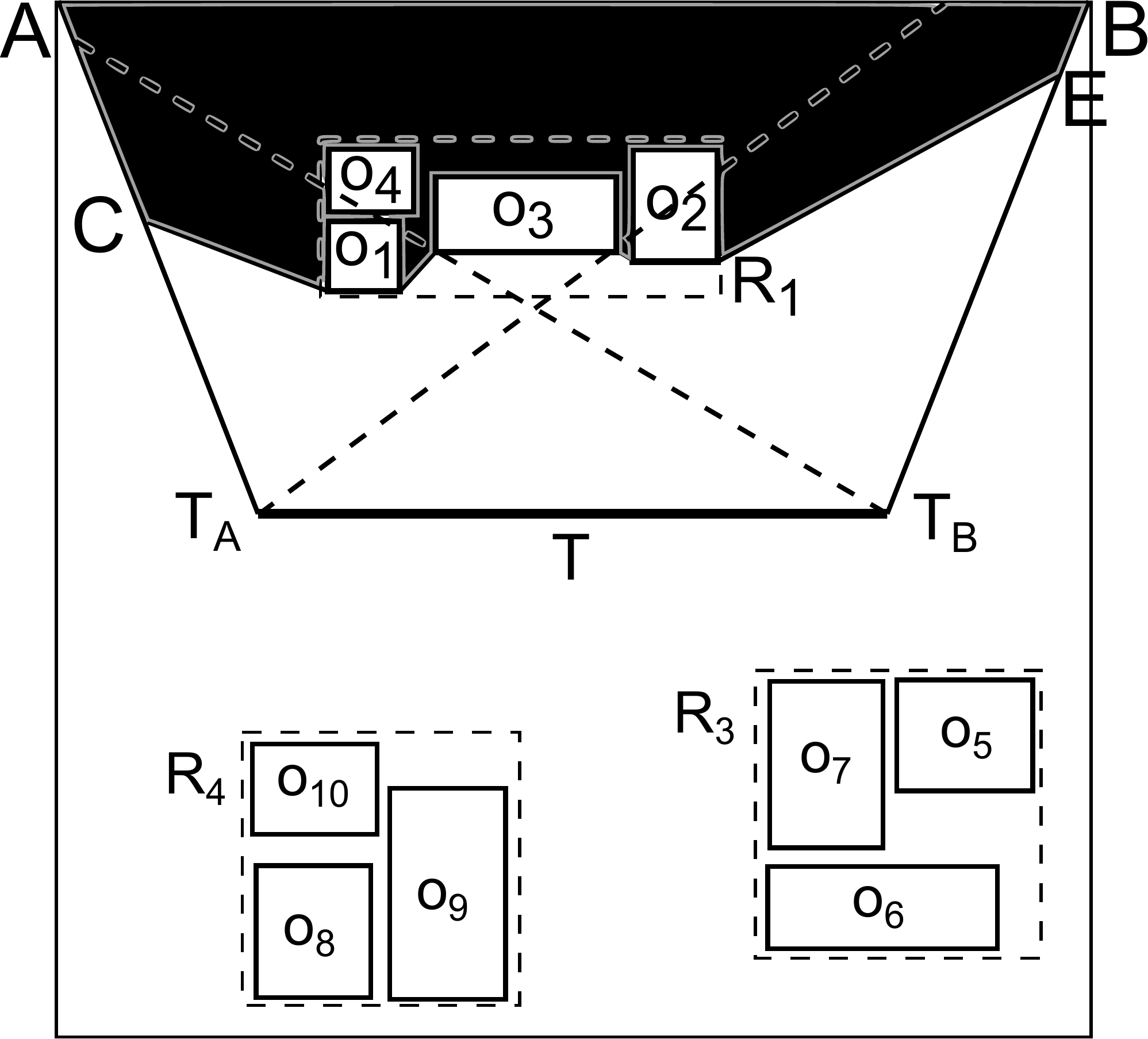}\label{fig:obs_region2}}
\hfill \caption{Visible region construction}
\vspace{-8mm}
\end{figure}

\eat{The work of Masud et al.~\cite{MOV} has shown how to compute visible regions for a given $T$ and $O$ from a specific viewpoint $q$~\cite{MOV}. This work maintains the set $S$ of visible regions to determine the obstacles that may further effect the visibility.
But in this case, the visible region is defined for a specific viewpoint and a target. In our case, we need to compute the visible region of the whole space with respect to a target where there is no specific viewpoint. So we maintain the visible region, but the computation of the visible region is different from Masud et al.}

A recent work~\cite{MOV} has developed a technique to determine the obstacles that affect the visibility of the target from a specific viewpoint. On the contrary, we need to compute the visible region of the whole space instead of a specific viewpoint with respect to a target. Thus, we cannot adopt the computation of the visible region from~\cite{MOV}. Our approach to compute the visible region in the presence of large number of obstacles is as follows.

Initially the set $V_R$ of visible regions contains the region of $R$ that is covered only by the field of view (FOV) with respect to the target $T$. As shown in Fig.~\ref{fig:obs_region1}, $V_R$ is the region bounded by the points $T_A$, $T_B$, $A$, and $B$ where $T_A$ and $T_B$ are the corner points of $T$. Here, FOV=$120^o$, the usual FOV of the human eye~\cite{FOV}. Initially, the set $O_R$ of obstructed regions is the region of $R$ that is outside the FOV. As $T$ cannot be viewed from the region outside the FOV, the obstacles residing in this region is discarded from consideration.

Next, we refine the set $V_R$ and the set $O_R$ by considering one obstacle at a time. The obstacle retrieval starts from the root node of the R*-tree. Only the nodes that intersect with a region in $V_R$ are incrementally retrieved from the R*-tree according to their non-decreasing distances from $T$. If the retrieved node is an MBR, its elements are further discovered. For example, in Fig.~\ref{fig:obs_region1}, when the MBR $R_1$ is accessed, its elements $o_1,o_2,o_3$, and $o_4$ are further discovered. If the retrieved node is an obstacle $o$, the regions in $V_R$ and $O_R$ are updated according to the effect of $o$.
We term the effect of a single obstacle $o$ on visibility as the shadow of $o$, $W_o$.

\begin{definition}
\emph{Shadow of an obstacle $o$, $W_o$.}
$W_o$ is the region formed by the set of points
$P$, where for any point $p$ of $P$, there is at least a point $t_p$ on $T$ such that the line segment joining $t_p$ and $p$ either
intersects or touches $o$.
\end{definition}

From each point $p$ of the $W_o$, $T$ is either completely or partially obstructed. The boundary of a shadow $W_o$ contains exactly two straight lines, which are tangents between the obstacle and the target. \eat{All the points of the $W_o$ lie to the left of one of these lines
in the direction from target to the obstacle. This line is defined as the left extension line. Presumably, the other line is the right extension line. }If these lines are rays, not the line segments that meet each other, then the region $W_o$ is unbounded. If $W_o$ is unbounded, we consider only the portion that is bounded by the given region $R$. In Fig.~\ref{fig:obs_region1}, The shadow of $o_1$ is $W_{o_1}$ (shown with black shade), the region bounded by $o_1$ and the points $A,C,$ and $D$.\eat{ Here, the left extension line is the line connecting $o_1$ with the point $D$ and the right extension line is the line connecting $o_1$ with $C$.}

While updating the $V_R$ and $O_R$ for the shadow of a retrieved obstacle, there are three cases to be considered:

(i) If the obstacle $o$ or its shadow $W_o$ does not overlap with any obstructed region of $O_R$, we exclude $W_o$ from $V_R$ and include it in $O_R$. In Fig.~\ref{fig:obs_region1}, obstacle $o_1$ is the first obstacle retrieved according to its non-decreasing distance from $T$. As there is no other obstructed region to be overlapped with $o_1$ or its shadow $W_{o_1}$ (shown as the black region), $W_{o_1}$ is now excluded from $V_R$ and included in $O_R$.

\eat{Similarly, the shadow $W_{o_2}$ of the next retrieved obstacle $o_2$ (shown with dotted lines) is also discarded from $S$ as $o_2$ does not overlap with any obstructed region.}

(ii) If $o$ or $W_o$ overlaps with one or more obstructed regions of $O_R$, we combine these regions and $W_o$ into a single obstructed region and discard this region from $V_R$. Let $N$ be the set of $W_o$ and the obstructed regions that overlap with $W_o$. To combine the regions in $N$, we determine the leftmost tangent line $l$ and the rightmost tangent line $r$ of all the shadows of $N$ such that the region bounded by $l$, $r$ and the union of the shadows of $N$ enclose all the obstructed regions of $N$. This shadow resembles the combined effect of the obstacles that are included in $N$. We replace the regions of $N$ from $O_R$ with this combined region and discard it from $V_R$. As an example, in Fig.~\ref{fig:obs_region1}, the shadow $W_{o_2}$ (shown with dotted lines) of the next retrieved obstacle $o_2$ overlaps with the existing obstructed region (shown with black shade). Here, the leftmost tangent line and the rightmost tangent line of these obstructed regions are the line connecting $o_1$, $C$ and the line connecting $o_2$, $E$, respectively. The region enclosed by these two lines and the union of the shadows is discarded from $V_R$.
Similarly, the next retrieved obstacle $o_3$ overlaps with $O_R$. The shadows combined for obstacle $o_1$, $o_2$, and $o_3$ is shown in black shade in Fig.~\ref{fig:obs_region2}, where the shadow of $o_3$ is shown with dotted lines.

(iii) If $o$ is entirely inside any obstructed region, it will not contribute to the visibility. So we discard $o$ from consideration. In Fig.~\ref{fig:obs_region2}, the effect of obstacle $o_4$ is not calculated as it is entirely inside the current obstructed region.

Note that, the visible region includes viewpoints from where the target is entirely visible. This approach is suitable for applications like placement of billboards, where partial visibility of the target from a viewpoint does not make sense. There can be some applications that require finding the viewpoints from where a target is partially visible. We leave computing the partial visible viewpoints as the scope of the future study.

\eat{As shown an example in Figure~\ref{fig:rtree}, $T$ is the
target and the obstacles $\{o_1,o_2,\dots,o_8\}$ are indexed in
three MBRs of an R*-tree, $\{R_1,R_2,R_3\}$ according to their
spatial orientation. These three MBRs are hierarchically organized
to form the root node. Initially the FOV is the visible region,
$V_i$. That is, $V_i$ equals the region bounded by the target $T$,
and points $A$ and $B$. Starting from the root node, $R_1,R_2,R_3$
are inserted in the priority queue $VQ$. Now the top of $VQ$ is
$R_2$, as it is the nearest node from $T$. But $R_2$ is completely
outside the visible region, therefore discarded from
consideration. The node $R_1$ is retrieved first as it is the next
nearest node from $T$ and it intersects with $ABT$. $R_1$ is an
MBR, so its elements $\{o_1,o_2,o_3\}$ are further discovered from
the R*-tree and inserted into $VQ$. Now $o_1$ is the top of $VQ$.
As $o_1$ is an obstacle and intersects with the visible region,
the obstructed region due to $o_1$ is computed as $x_1x_2x_3$. The
visible region is now the region $V_i - x_1x_2x_3$. This procedure
continues according to the above mentioned process. The polygonal
region bounded by dotted lines in Figure~\ref{fig:rtree} is the
obstructed region obtained after the visible region construction
phase.}

\subsection{Merging Cells with Visible Regions}\label{sec:merging}
In the final step of producing a VCM, we combine colors computed in the first step (that ignores obstacles) and visible regions computed in the second step to compute color of each cell in the presence of obstacles. Intuitively, all obstructed regions are assigned color zero (representing zero visibility), while all visible regions are assigned colors from their corresponding cells.
This requires taking intersection (spatial join) of all cells and visible regions.

The above process can be expensive for two reasons. First, visible
regions are usually irregular polygons and intersecting them with
cells incurs high computational overhead due to the complex shapes of
the polygons. We address this by storing each polygon as a set of
regular shapes (rectangles) with a quad-tree~\cite{quadtree}.
Second, the number of cells can be quite large and intersecting
visible regions with all cells can be expensive. We address this
by indexing the cells in an R*-tree and intersecting each visible
region only with its overlapping cells. We describe these
optimizations below.

\subsubsection{Indexing Visible Regions}
\label{sec:vrqt}

We index visible (and obstructed) regions in a 2D (3D) space with a quad-tree (octree). Quad-trees (Octrees) partition a 2D (3D) space by recursively subdividing it into four quadrants (eight octants) or blocks. Initially the whole space is represented with a single quad-tree block. In the visible region computation phase, when a region is obstructed due to a retrieved obstacle, a quad-tree block is partitioned into four equal blocks if it intersects with the obstructed region.
The partitioning continues until (i) a quad-tree block is
completely visible, or (ii) completely obstructed, or (iii) the
size of a block is below a threshold value $\vartheta$. The
threshold size is determined from the partitioning phase. If the
set of the boundary points of an equivalence cell $\zeta_i$ is
$\beta_{\zeta_i}$, the minimum distance between any two opposite
boundaries over all cells is specified as $\vartheta$, i.e.,

\vspace{-4.8mm}
$$
\vartheta = \argmin_{\zeta_i, 1 \le i \le n}(\argmin_{p,p^\prime
\in \beta_{\zeta_i}} (mindist(p,p^\prime))
\vspace{-3.6mm}
$$
where,  points $p,p^\prime \in \beta_{\zeta_i}$ are points of
opposite boundaries.

As $\vartheta$ is the minimum size of a cell obtained in the partitioning phase and the deviation in the perceived visibility for a block-size less than $\vartheta$ is not distinguishable due to angular resolution of a lens, dividing a quad-tree block into size smaller than $\vartheta$ for more accuracy is redundant.

\begin{figure}
\label{fig:quadtrees_VCM}
\centering \subfloat[]{\includegraphics[width=1.65in]{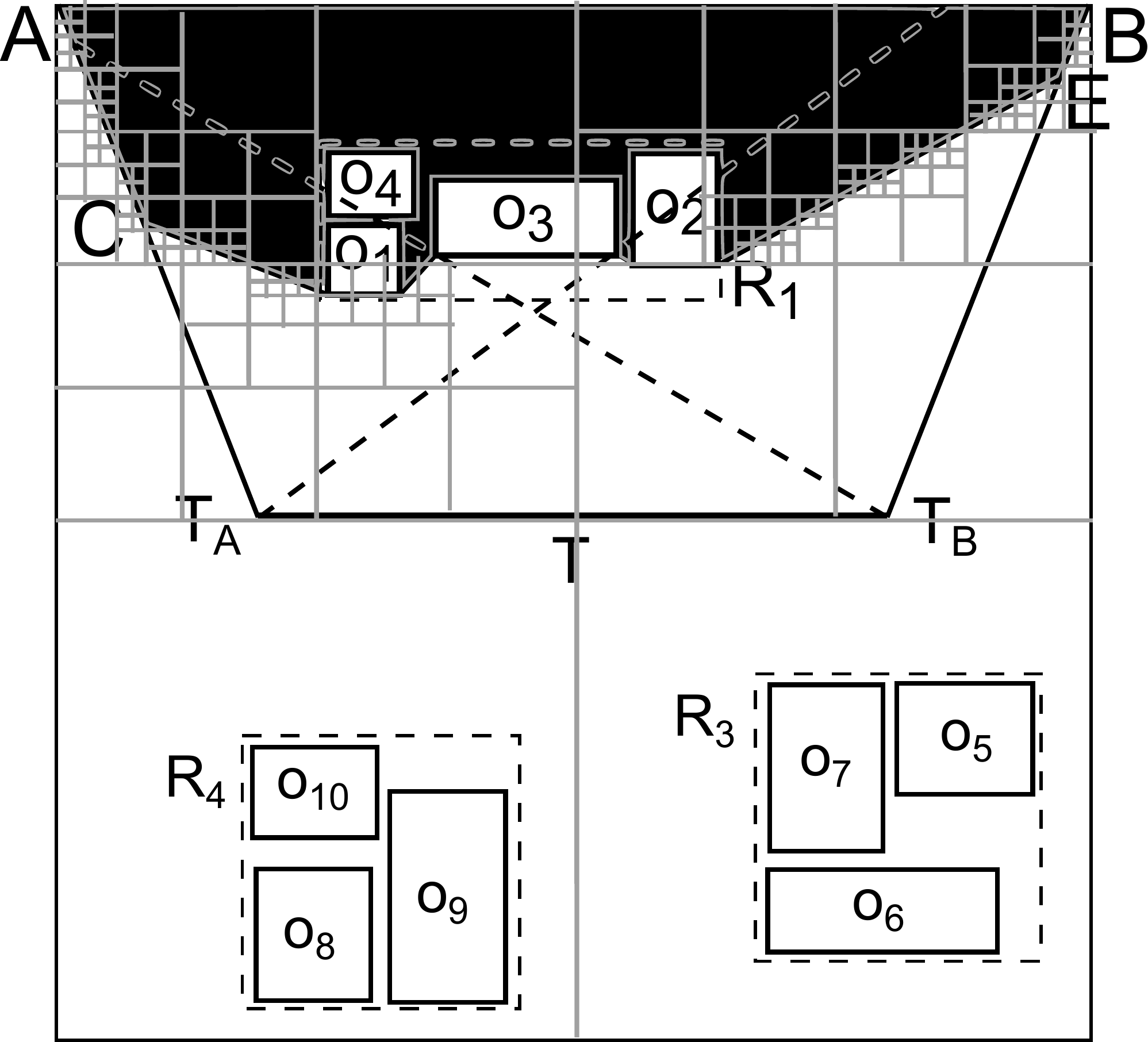}\label{fig:quad}}
\hfill \subfloat[]{\includegraphics[width=1.65in]{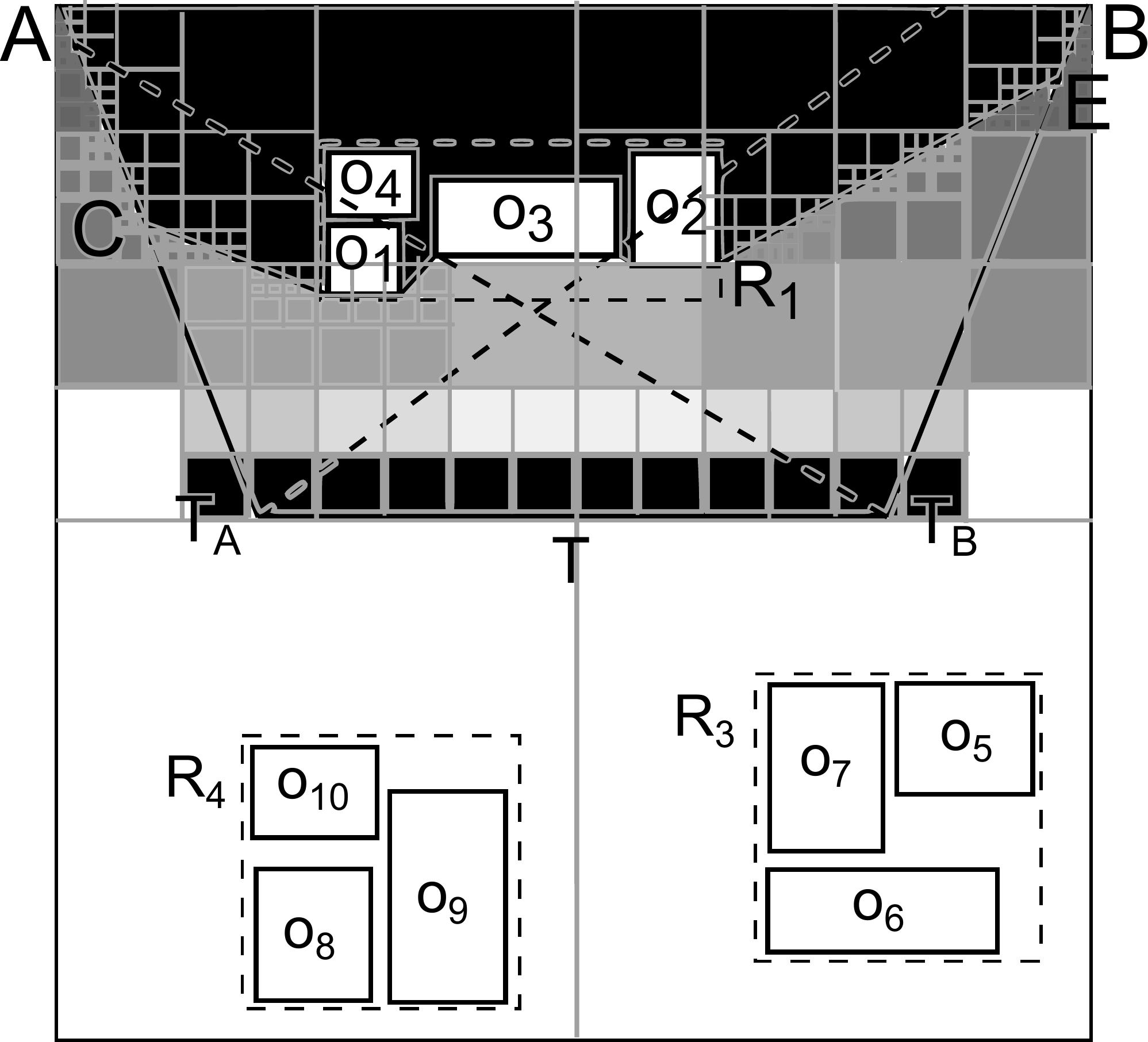}\label{fig:VCM}}
\hfill \caption{Construction of visibility color map}
\vspace{-8mm}
\end{figure}

Fig.~\ref{fig:quad} shows the quad-tree of the obstructed region (black blocks) and visible region (white blocks inside FOV).
After the quad-tree is constructed, we start assigning color to each of its blocks. We get the VCM after all the blocks are assigned colors. The blocks that fall in the obstructed region are assigned zero visibility. The remaining blocks (i.e., visible region) are assigned colors based on the colors of their corresponding cells, as described next.
%
%

\eat{Note that, in many practical applications, there are usually
many obstacles present in a space. So most of the quad-tree blocks
are likely to be obstructed in the visibility region construction
phase. Therefore we represent the quad-tree blocks as a linear
list data structure. XXX clarify XXX}


\subsubsection{Indexing Cells and Constructing VCM}

For each quad-tree block with unassigned color, we need to find
all overlapping cells in order to find its color. To expedite this
process, we index cells with an R*-tree that we call the
\emph{color-tree}. Leaf nodes in the color-tree represent cells and
non-leaf nodes represent MBRs containing the shapes of their
children nodes. Then, for each quad-tree block, we run a range
query on the color-tree. The color for an unassigned quad-tree
block is obtained by calculating the intersection of the spatial
region of that block and the cells from the color-tree. If the
quad-tree block intersects with a single cell of the color
partition, the block is assigned the color of that cell. If a
quad-tree block intersects with multiple cells, in a 2D (3D) space
we further divide that block into four (eight) equal blocks. The
division is continued until either a block intersects with a
single cell or the size of a block is below the threshold value
$\vartheta$ (Section~\ref{sec:vrqt}). The process
terminates when all quad-tree blocks of the visible region are
colored according to the visibility measure.
Fig.~\ref{fig:VCM} shows the resulting VCM constructed
by combining the color partitions of Fig.~\ref{fig:partition}
and the quad-tree of Fig.~\ref{fig:quad}.

The steps of constructing a VCM are shown in
Algorithm~\ref{algo:construct}. Lines 1.6-1.9 shows accessing the
color-tree nodes. If an accessed node is an MBR, its elements are
further discovered from the color-tree (Lines
1.21-1.22). If the accessed node is a leaf node, the quad-tree
nodes that are not colored yet and intersects with this node are
either colored or partitioned further (Lines 1.15-1.20). Finally, the colored quad-tree is returned as
the complete VCM.

\vspace{-3mm}
\setlength{\algomargin}{1.2em}
\begin{algorithm}[h]
\caption{\emph{ConstructVCM($T, FOV,  \vartheta, Qtree, CTree$)}}
\label{algo:construct}
\begin{small}
Initialize $Q$ to an empty queue\; Initialize $L$ to an empty
list\; $node \leftarrow {CTree.root}$; $end \leftarrow$ false\;

$L \leftarrow Get\_unassigned\_quad_\_leaf(Qtree,FOV)$\;

\While {$node \neq$ empty and $end$ = false} {
    \For{each element $n_e$ of $node$}
    {
        \uIf{Inside\_visibility\_region($n_e$) = true}
        {
            $Enqueue(Q,n_e)$\;
        }
    }

    $continue \leftarrow$ true\;
    \While {$continue$ = true}
    {
        $continue \leftarrow$ false;  $node \leftarrow Dequeue(Q)$\;
        \uIf{$node$ = empty \textbf{or} $L$ = empty}
        {
            $end \leftarrow$ true\;
        }
        \uElseIf {$node$ is a data object}
        {
            Initialize $Still\_unassigned\_quads$ to an empty list\;
            \For{each element $L_i$ of $L$}
            {
                $Still\_unassigned\_quads \leftarrow insert(Divide\_and\_color(L_i,node,\vartheta))$\;
            }
            $L \leftarrow Still\_unassigned\_quads$\;

            $continue \leftarrow$ true\;
        }
        \Else
        {
            $node \leftarrow$ child($node$)\;
        }
    }
}
\eat{Initialize $L$ to an empty list\; $L \leftarrow$
Partitioning($T, \mu, N_e$)\;

\For{each element $L_e$ of $L$} {
    \uIf{InsideFOV($q_e$) = true
}

\For{each element $q_e$ of $Qtree$} {
    \uIf{InsideFOV($q_e$) = true \textbf{and} Colored($q_e$) = false}
    {
        $P \leftarrow Intersection(q_e, L)$\;
        \uIf{lengthof($P$) = $1$}
        {
            $Colorof(q_e) \leftarrow Colorof(P)$\;
            $Colored(q_e) \leftarrow true$\;
        }
        \Else
        {
            $D \leftarrow Divide(q_e,\epsilon$)\;
            \For{each element $d_e$ of $D$}
            {
                $P \leftarrow Intersection(d_e, L)$\;
                $Colorof(d_e) \leftarrow Colorof(P)$\;
                $Colored(d_e) \leftarrow true$\;
            }
        }
    }
} } } return $Qtree$;
\end{small}
\end{algorithm}
\vspace{-5mm}


\eat{To construct the visibility color map  we need to compute the
intersection of the color-tree blocks with the blocks of the
quad-tree. To improve efficiency in this process, we have
introduced three approximations for the color-tree partitions and
compared their performances.}

Since the color partition results in complex shaped cells
(e.g., curves), it is computationally expensive to combine these
shapes with the quad-tree blocks of visible region. Thus,
we propose two approaches to approximate color
partition cells.

\section{Extensions}
In this section, we discuss a few extensions to our basic algorithm described in the previous section.

\subsection{Approximation of Partitions}\label{sec:approx}
The algorithm in the previous section partitions the space according to the relative
distance and angle between the lens and the target. Based on these
parameters, the cells are bounded by arcs and straight lines. To
construct the VCM, we need to compute the
intersection of these cells with the quad-tree blocks. The process
is computationally expensive due to the complex shape of the
cells and the target.

To address this, we introduce two approximations that reduce the computational overhead at the cost of small bounded errors:
(i) minimum bounding rectangle (MBR) of a cell and (ii) tangents
of the arcs of a cell. For the
ease of explanation we analyze errors for targets with regular
shapes, e.g., lines without loss of generality. For such targets, the complex shaped
partitions are bounded by two arcs and two straight lines. Here we
discuss the approximations of color partitions and the maximum
error resulting from these approximations.

\begin{figure}
\centering
\subfloat[]{\includegraphics[height=1in]{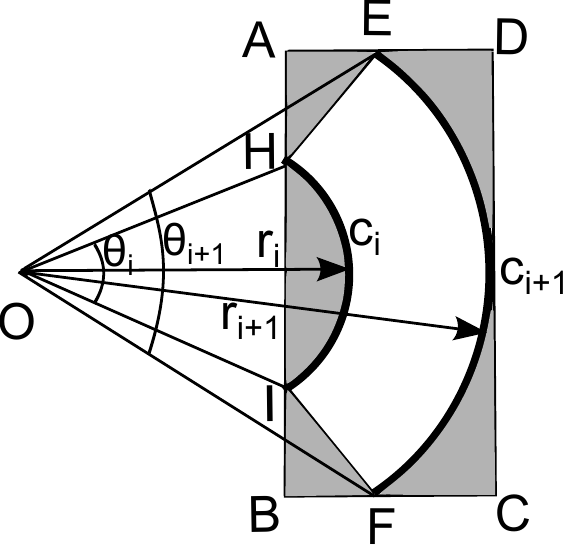}\label{subfig:a}} \hfill
\subfloat[]{\includegraphics[height=1in]{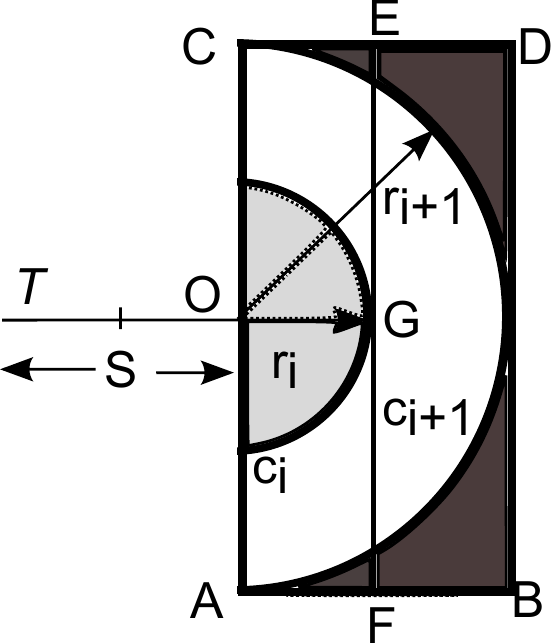}\label{subfig:b}} \hfill
\subfloat[]{\includegraphics[height=1in]{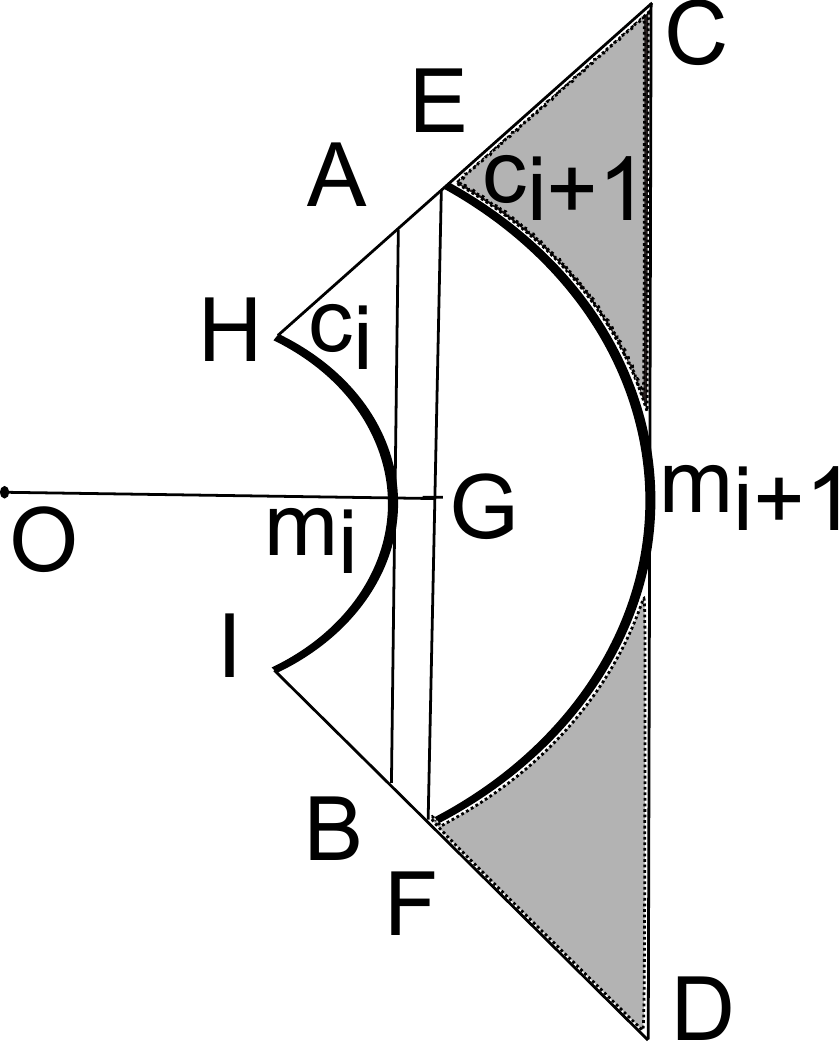}\label{subfig:c}}
\caption{Approximating a cell using MBR and tangents}
\vspace{-7mm} \label{fig:MTapproximation}
\end{figure}

\subsubsection*{MBR Approximation}

An approach to approximate the curves of a cell is to enclose the
cell using its MBR where the area covered by the cell is
approximated by the area covered by the enclosing MBR. This is
illustrated in Fig.~\ref{subfig:a}, where a cell
consists of two concentric arcs $c_i$ and $c_{i+1}$ (centered at a
corner $O$ of the target) of radius $r_i$ and $r_{i+1}$,
respectively. $ABCD$ is the MBR of this cell. Let $c_i$ and
$c_{i+1}$ create $\theta_i$ and $\theta_{i+1}$ angles (in degrees)
at the center respectively. We denote the area of the MBR as
$A_r$. So the area bounded by the cell, $A_p$ is
$(\frac{\theta_{i+1}}{360} \times \pi{r_{i+1}}^2)-
(\frac{\theta_i}{360} \times \pi{r_i}^2)$. Hence the area that
gets wrong color due to this MBR approximation is $A_r- A_p$ (shaded region in Fig.~\ref{subfig:a}).

\noindent{\bf Error Bound Analysis.}
For targets with regular shapes such as lines
(Fig.~\ref{fig:partition}), the largest cell that can yield the
maximum possible error consists of two half circles centered at a
corner point of the target. Hence we formulate the maximum error
bound by referring to Fig.~\ref{subfig:b}. Here,
we want to approximate the area of a cell $p_i$ bounded by two
half circles $c_i$ and $c_{i+1}$ centered at $O$, a corner point
of target $T$ of length $S$. Here, $c_i$ and $c_{i+1}$ belong to
distance based partition $d_i$ and $d_{i+1}$, respectively. We
approximate $p_i$ by taking its MBR, $ABCD$. So, the darkly shaded region is wrongly colored for MBR
approximation. The
lightly shaded region is wrongly colored too, but it is considered
for distance based partition $<d_{i-1},d_i>$ i.e., for cell
$p_{i-1}$. As the total error is calculated incrementally for each
cell, so the error for this lightly shaded region is calculated
only once for the cell $p_{i-1}$. Let, $r_i$ and $r_{i+1}$ be the
radii of $c_i$ and $c_{i+1}$, respectively. Hence, the width of
cell $p_i$ is $r_{i+1} - r_i$. The farthest distance from $O$ to
any point of the MBR is $\Delta r = \sqrt{2} \times r_{i+1}$. So
the points of the MBR that fall within the distance $r_{i+1}$ and
$\sqrt{2} \times r_{i+1}$ from point $O$, get wrongly colored with
the $p_i$'s color. From Equation~\ref{eq} we get that the width of
the distance based partitions increase with the increase of the
distance between the target and the partition. So the maximum
number of distance based partitions in $d_{i+1}$ that can be
wrongly colored is $n =\frac{\Delta r - r_{i+1}}{r_{i+1}-r_i}$.

For each distance based partition, corresponding angle based
partitions are calculated by taking the angle from the midpoint of
the target. So only the angle based partitions that fall within
the range ($\arctan \frac{r_{i+1}}{S/2}$, $-\arctan
\frac{r_{i+1}}{S/2}$) can lie in the darkly shaded region.
Let there are $a_i$ such angle based partitions in total. As the
difference in visual angle in consecutive partitions is $\mu$, the
maximum variation in color for a cell is, $\Delta c_{m_i} = n
\times a_i \times \mu$. So when the total number of partitions is
$k$ and the area of $i$th cell is $A_i$, the total error in
coloring is:
\vspace{-2mm}
\begin{equation}
\label{eq_np}
E_{MBR} = \sum_{i=1}^{k} \Delta c_{m_i} \times A_i
\end{equation}
\vspace{-4mm}

\subsubsection*{Tangential Approximation}

We obtain another approximation approach by taking tangents in the
midpoints of the two arcs that encloses a cell. Here the area
enclosed by two arcs is approximated by the area enclosed by the
tangents at their midpoints. In
Fig.~\ref{subfig:c} the cell is bounded by arcs
$c_i$ and $c_{i+1}$ centered at a corner $O$ of the target $T$. We
take two tangents $AB$ and $CD$ of $c_i$ and $c_{i+1}$ at their
midpoints $m_i$ and $m_{i+1}$, respectively. So we want to
approximate the area bounded by these two arcs with the trapezoid
$ABDC$. Let, the angles created at $O$ by $c_i$ and $c_{i+1}$ are
$\theta_i$ and $\theta_{i+1}$ (in degrees). If we color the
trapezoid $ABDC$ instead of the cell, the region that gets wrong
color is the shaded region of Fig.~\ref{subfig:c}.
Let the area of the trapezoid $ABDC$ be $A_{t_1}$, the radius of
$c_{i+1}$ is $r_{i+1}$, the midpoint of $\overline{EF}$ is $G$,
and the length of $\overline{OG}$ is $x$. Then the length of
$\overline{EG}$ is $\sqrt{{r_{i+1}}^2- x^2}$. The area of the
segment bounded by $\overline{EG}$ and $c_{i+1}$, $A_c$ is
$(\frac{\theta_{i+1}}{360} \times \pi{r_{i+1}}^2)-x \times
\sqrt{{r_{i+1}}^2- x^2}$. If the area of the trapezoid $ABFE$ is
$A_{t_2}$, then the area of the shaded region $A_e$ is $A_{t_1}-
A_c-A_{t_2}$. According to this tangential approximation the
region that is wrongly colored due to cell $HIFE$ is this shaded
region with area $A_e$.

\noindent{\bf Error Bound Analysis.}
\eat{In general, for target shapes similar to
Fig.~\ref{fig:partition}, the largest cell that can result into
maximum possible error consists of two half circles. }In case of
MBR approximation the cell bounded by two half circles is
approximated by an enclosing MBR, while for tangential
approximation that cell is approximated by a rectangle bounded by
the tangents of those two half circles. Referring to
Fig.~\ref{subfig:b}, the cell bounded by arcs
$c_i$ and $c_{i+1}$ is approximated by rectangle $ABDC$ and
rectangle $FBDE$ for MBR and tangential approaches, respectively. In
this figure, $G$ is the midpoint of $c_i$ and the farthest
distance from $G$ to any point inside the approximated cell is
$\Delta r = \sqrt{r_{i+1}^2+ (r_{i+1}-r_i)^2}$. So the points
inside the darkly shaded region are wrongly colored. As the width of the distance
based partitions relates inversely to the distance between the
target and the partition (Equation~\ref{eq}), the maximum number
of distance based partitions in $d_{i+1}$ that can be wrongly
colored is $n =\frac{\Delta r - r_{i+1}}{r_{i+1}-r_i}$. In case of
angle based partitioning, only the angle based partitions that
fall within the range ($\arctan \frac{r_{i+1}}{S/2+r_i}$,
$-\arctan \frac{r_{i+1}}{S/2+r_i}$) can lie in the darkly
shaded region inside the rectangle $EFBD$. Let there are $a_i$ such angle based partitions in
total. As the difference in visual angle in consecutive partitions
is $\mu$, the maximum variation in color for a cell is, $\Delta
c_{t_i} = n \times a_i \times \mu$. So when the total number of
partitions is $k$ and the area of $i$th cell is $A_i$, the total
error is:
\vspace{-2mm}
\begin{equation}
\label{eq_np}
E_{Tangent} = \sum_{i=1}^{k} \Delta c_{t_i} \times A_i
\end{equation}
\vspace{-2mm}

\eat{ Due to this similarity in approximated cell shape, the
analysis of maximum error is similar for both tangential and MBR
approximations. Hence the total error for coloring using
tangential approximation directly follows from Equation
~\ref{eq_np}.}

In the above sections, we discussed our approach to construct
\emph{target-centric visibility color map}. The approach is same
for the \emph{viewer-centric visibility color map} with an
additional initialization step. The details of
\emph{viewer-centric VCM} is discussed below.

\subsection{Viewer-centric VCM}
\label{sec:viewer} A \emph{viewer-centric} VCM is constructed by
calculating the visibility of the surrounding space for a given
viewpoint $q$ and a set $O$ of obstacles. Unlike the
\emph{target-centric} VCM where a particular target is specified,
in case of the \emph{viewer-centric} VCM, a particular viewer
position is specified\eat{The difference between the
\emph{viewer-centric} VCM and the \emph{target-centric} VCM is, a
particular target is explicitly specified instead of a viewpoint
in the latter case}. To measure the visibility using
Equation~\ref{eq} and Equation~\ref{eq2} in case of the
viewer-centric VCM, first we need an initial value of the size $S$
of a target from the given information.

The visibility of any point farther than a distance $d_{max}$ is
not visually distinguishable if the perceived visual angle of that
point is less than the angular resolution $\mu$ from $q$ (Section~\ref{subsubsec:parition}). Based on
this fact, we assume a circular region $R$ of radius $d_{max}$
centered at $q$ to construct the VCM of $R$ only, as the
visibility of the points outside $R$ are not distinguishable by a
viewer at $q$. Using Equation~\ref{eq}, we calculate the size $S$
of a target for which visual angle $V = \mu$ is perceived at
distance $d_{max}$. After taking $S$ as the size of the target,
the steps of our approach discussed for the \emph{target-centric}
VCM is applied in the same manner to construct the
\emph{viewer-centric} VCM.

\subsection{Incremental Processing}
\label{sec:IP} During the computation of the VCM, we have assumed that
the field of view (FOV), i.e., the orientation of the viewpoint or
the target is fixed. At a particular orientation or gaze
direction, only the extents of the space that is inside the FOV,
is visible. With the change in viewing direction, some areas that
were previously outside the FOV become visible. In this case, we do
not need to compute the full VCM each time, rather we can
incrementally construct the VCM by computing for the newly visible
parts only.

In the incremental process, the only information that varies is
the gaze direction. So, as a preprocessing step we can construct
the color-tree and the visible region quad-tree in the above
discussed method by considering the FOV as $360^o$. A VCM is then
constructed by combining the color-tree and the quad-tree for a
particular gaze direction. When the gaze direction changes, only
the uncolored quad-tree blocks that are now included in the
visibility region are assigned colors. Hence the color-tree and
the quad-tree are constructed only once. This reduces the
computational complexity to a great extent by avoiding same
calculations repetitively. From our conducted experiments we also
observe that the processing time required to combine the
color-tree and the quad-tree is much smaller than the processing time
required to construct the visibility region quad-tree for a dataset
of densely distributed large number of obstacles (Section~\ref{subsec:ar_variation} and ~\ref{subsec:3dar_variation}). So for such cases we can significantly improve the performance of our proposed
solution by adopting the preprocessing strategy.

\section{Experimental Evaluation}
\label{sec:exp} We evaluate the performance of our proposed
algorithm for constructing the \emph{visibility color map} (VCM)
with two real datasets. Specifically, at first we compare our approach with a \emph{baseline approach} that approximates the total space into a regular grid and compute visibility from the midpoint of each grid cell, and then we compare our two
approximation algorithms, i.e., \emph{MBR approximation} ($VCM_M$) and
\emph{tangential approximation} ($VCM_T$) with the exact method
($VCM_E$). The algorithms are implemented in C++ and the
experiments are conducted on a core i5 2.40 GHz PC with 3GB RAM,
running Microsoft Windows 7.

\subsection{Experimental Setup}
\label{subsec:setup} Our experiments are based on two real
datasets: (1)
British\footnote{\url{http://www.citygml.org/index.php?id=1539}}
representing 5985 data objects obtained from British ordnance
survey\footnote{\url{http://www.ordnancesurvey.co.uk/oswebsite/indexA.html}}
and (2)
Boston\footnote{\url{http://www.bostonredevelopmentauthority.org/BRA_3D_Models/3D-download.html}}
representing 130,043 data objects in Boston downtown\eat{ as depicted
in Table~\ref{table:real_dataset}}. In both datasets, objects are
represented as 3D rectangles that are used as obstacles in our
experiments. For both datasets, we normalize the dataspace into a
span of $10,000\times10,000$ square units. For 2D, the
datasets are normalized by considering $z$-axis value as 0.
All obstacles are indexed by an R*-tree, with the disk page size
fixed at 1KB.

\eat{\begin{table}
\centering \caption{Description of datasets} \vspace{-8pt}
\label{table:real_dataset}
\begin{small}
\begin{tabular}{|l|l|l|}
\hline
{\bfseries Dataset} & {\bfseries Cardinality} & {\bfseries Description}\\
\hline
British & 5985 & British Ordnance Survey\\
\hline
Boston & 130043 & Boston Downtown\\
\hline
\end{tabular}
\end{small}
\vspace{-10pt}
\end{table}

}

\begin{table}
\centering \caption{Parameters} \label{table:param}
\begin{small}
\begin{tabular}{|p{3.24cm}|l|l|}
\hline
{\bfseries Parameter} & {\bfseries Range} & {\bfseries Default}\\
\hline
Angular Resolution ($\mu$) & 1, 2, 4, 8, 16 & 4\\
\hline
Minimum Block Size ($\vartheta$) & 1, 2, 4, 8, 16 & 1\\
\hline
Query Space Area ($A_Q$) & 0.05, 0.10, 0.15, 0.2, 0.25 & 0.15 \\
\hline
Field of View ($FOV$) & 60, 120, 180, 240, 300, 360 & 120 \\
\hline
Length of Target ($L_T$) & 0.05, 0.10, 0.15, 0.2, 0.25 & 0.15 \\
\hline
Dataset Size ($D_S$) & 5k, 10k, 15k, 20k, 25k & \\
\hline
\end{tabular}
\end{small}
\vspace{-16pt}
\end{table}

The experiments investigate the performance of the proposed
solutions by varying five parameters: (i) angular resolution
($\mu$) in arcminutes, (ii) the threshold of the quad-tree block
size ($\vartheta$) as the multiple of the calculated minimum size
(as explained in Section~\ref{sec:vrqt}), (iii) the area of the
space ($A_Q$) as the percentage of the total area, (iv) field of
view ($FOV$) in degrees, and (v) the length of the target ($L_T$)
as the percentage of the length of total dataspace. We have also varied the dataset size using both \emph{Uniform} and \emph{Zipf} distribution of the obstacles. The range and default
value of each parameter are listed in
Table~\ref{table:param}. In concordance with human vision, the
default values of $\mu$ and $FOV$ are set as $4$ arcminutes~\cite{angResolutionHuman} and
$120^o$~\cite{FOV}, respectively. The minimum
threshold of quad-tree block size as calculated in
Section~\ref{sec:vrqt} is used as the default value of
$\vartheta$. The default values of other parameters are set to
their median values.  \eat{In each experiment we vary only one
parameter and set the other parameters at their default values.}

The performance metrics that are used in our experiments are: (i)
the total processing time, (ii) the total I/O cost, and (iii) the
error introduced by the two approximations: $VCM_M$ and $VCM_T$.
We calculate the approximation error as the deviation from the
color map of $VCM_E$, i.e.,
\vspace{-2mm}
\begin{equation}
\label{error_eq} error= \frac{\sum_i (c_{e_i} \times A_i - c_{a_i}
\times A_i)}{\sum_i c_{e_i} \times A_i}
\end{equation}

Here $c_{e_i}$ is the color of $i$th cell in $VCM_E$, $c_{a_i}$ is
the color of $i$th cell in $VCM_M$ or $VCM_T$, and $A_i$ is the
area of the $i$th cell.

For each experiment, we have evaluated our solution for the target
at 100 random positions and reported their average performance. We have conducted extensive experiments using two datasets for both 2D and 3D spaces. Since, 2D dataspace is a subset of 3D dataspace and most of the real applications involve 3D scenario, we omit the detailed results of 2D due to the brevity of presentation.

\eat{In Section~\ref{subsec:straightforward}, we compare the performance
of our solution with the baseline approach. Subsequently in
and Section~\ref{subsec:3D}, we compare
the performance of our proposed solutions in 2D and 3D spaces.}


\begin{figure}
\label{fig:dataset2} \centering \subfloat[British ordnance
survey]{\includegraphics[height=0.7in]{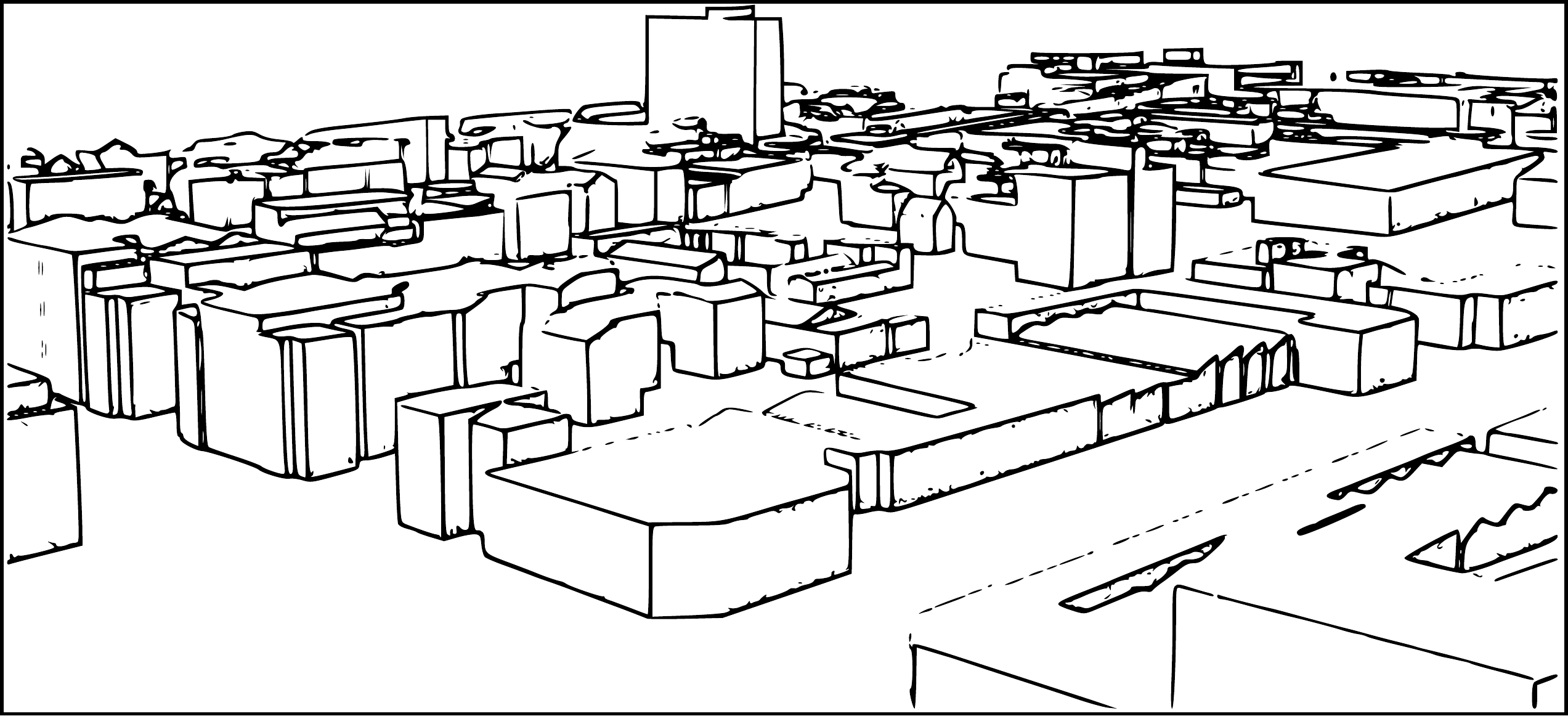}}\hfill
\subfloat[Boston
downtown]{\includegraphics[height=0.7in]{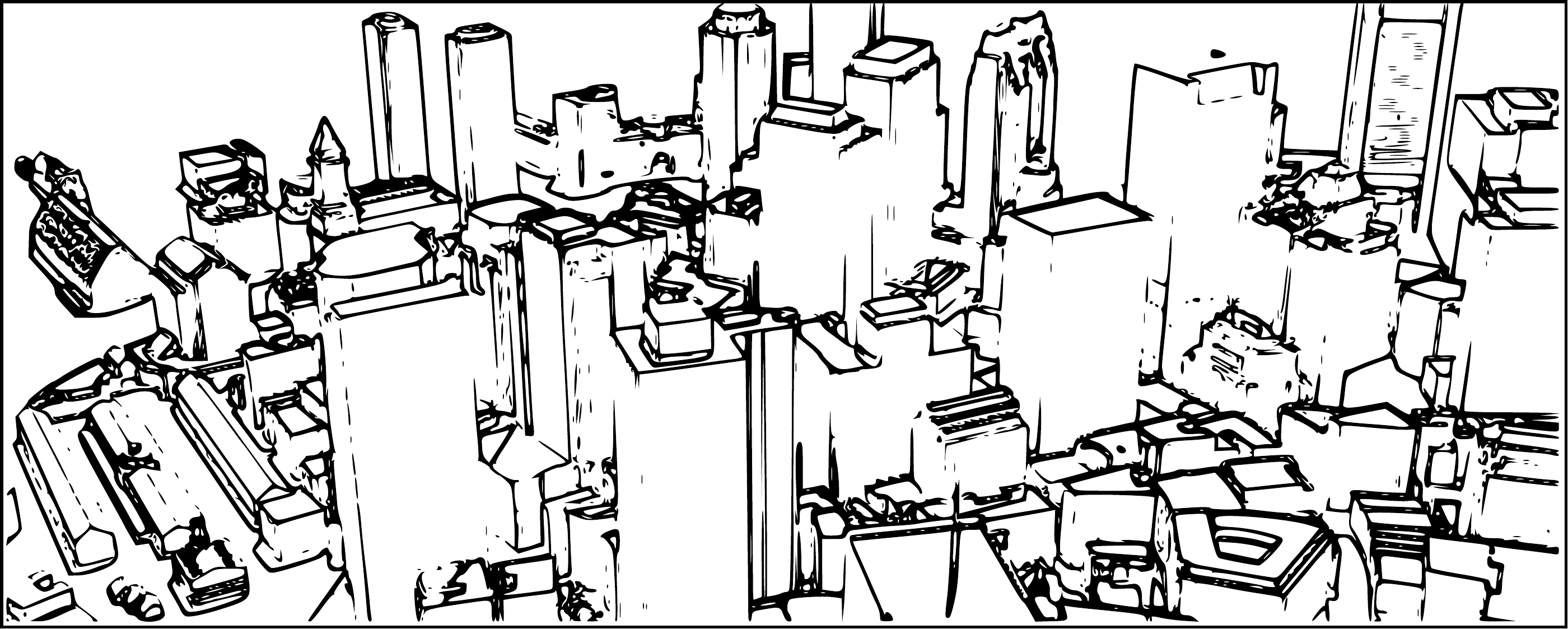}}
\caption{Real datasets used in the experiments} \vspace{-8mm}
\end{figure}

\begin{figure}[h]
\label{fig:dataset1}
\centering
\vspace{-5mm}
\subfloat[British ordnance
survey]{\includegraphics[height=1.1in]{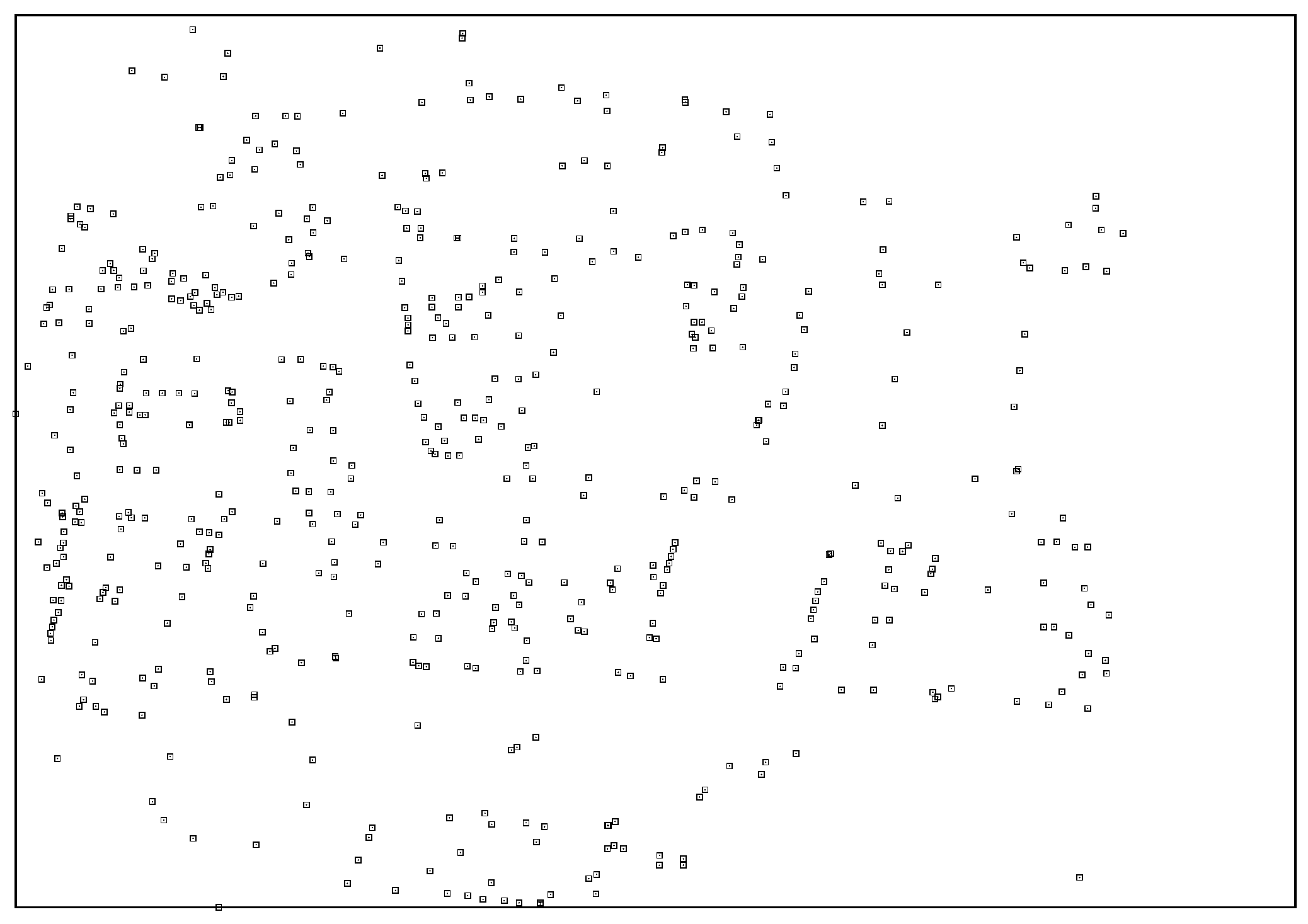}}\hfill
\subfloat[Boston
downtown]{\includegraphics[height=1.1in]{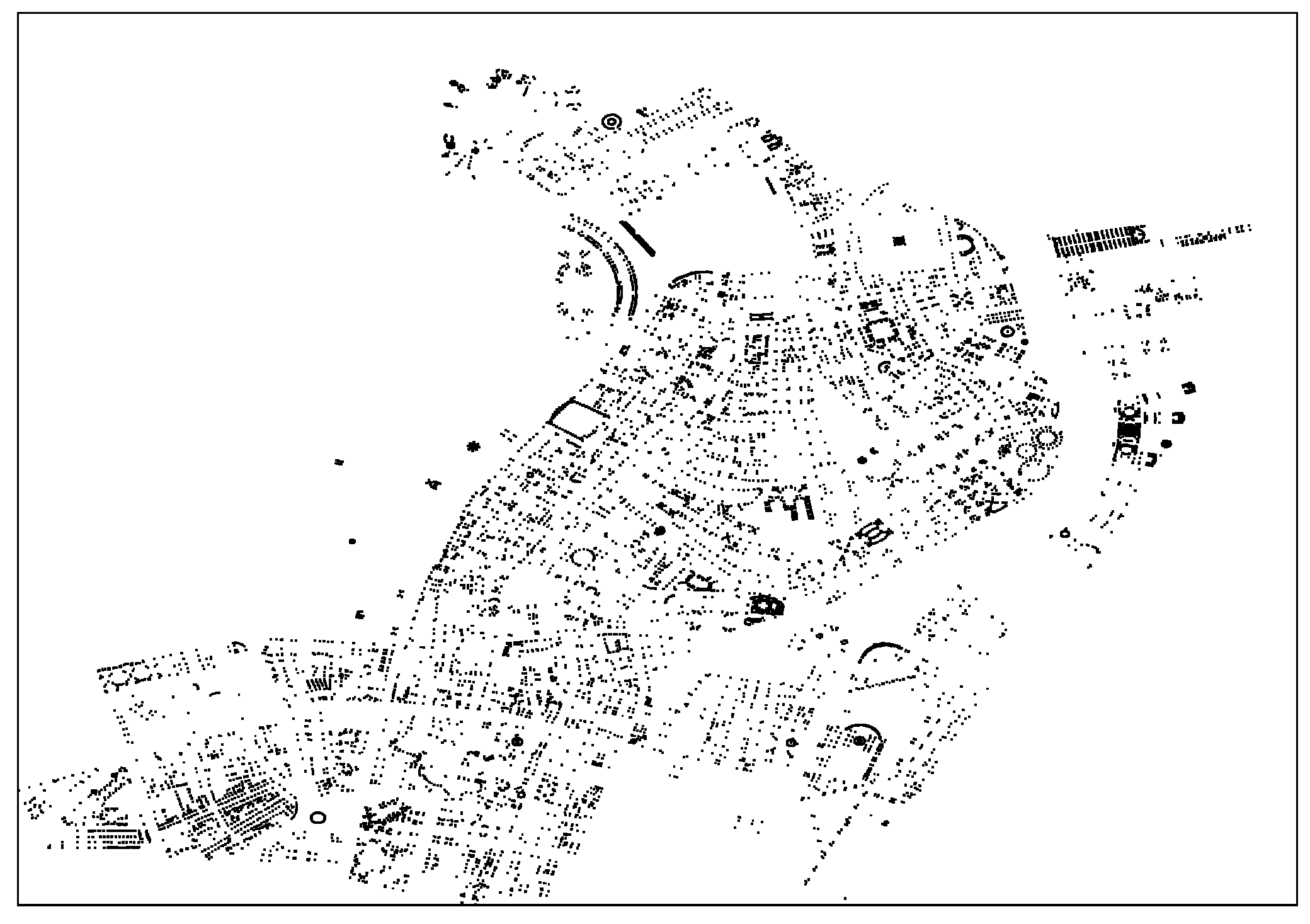}}
\caption{Dataset distribution}
\vspace{-6mm}
\end{figure}

\subsection{Comparison with the Baseline}
\label{subsec:straightforward}
The performance improvement of our
approach over the \emph{baseline} approach is measured in terms
of the total processing time required for computing the VCM. In the na\"{\i}ve approach, we need to compute visibility
of the target from \emph{infinite} number of points to construct
the VCM as every point in the dataspace acts as a viewpoint. Even
discretizing the surrounding space into 1000 points in each dimension
would give a total of $10^9$ points in the 3D space.
We observe that the time required for
such computation using British and Boston datasets are
approximately 74 days and 128 days, respectively. As such na\"{\i}ve approach is trivial, we can further improve it by dividing the dataspace into $500\times500\times500$ cubic cells in 3D and choose the visibility from the middle point of each cell to \emph{represent} the visibility of that entire cell. In the rest of the paper we refer to this approach as the \emph{baseline} approach to construct the VCM. The comparison between the baseline approach and our proposed solution $VCM_E$ is presented in Table~\ref{table:baseline} in terms of total processing time and introduced error. In this case the parameters are set to their default values (Table~\ref{table:param}). We observe that when compared to $VCM_E$, for British dataset the baseline approach runs 818 times slower and introduces $30$\% error, while for Boston dataset the baseline approach runs 837 times slower and introduces $32$\% error.
%

\eat{On the other hand, the performance of our proposed solutions
is measured by the time required in three major steps of a VCM
computation: partition the space based on visibility measure,
construct visibility region quad-tree by retrieving obstacles from
the database, and finally, combine the outcome of previous two
steps to produce the VCM.Our solution outperforms the na\"{\i}ve
approach by at least \emph{four orders} of magnitude. This huge
improvement comes from our novel partitioning and visibility
region computation techniques that discard the visibility
computation for a large number of viewpoints and retrieve only
necessary obstacles from the database.}

\begin{table}[h]
\centering
\vspace{-2mm}
\caption{Comparison with the baseline}
\label{table:baseline}
\begin{small}
\begin{tabular}{|l|l|l|l|}
\hline
Dataset & \multicolumn{2}{|c|}{Baseline} & $VCM_E$ \\
\cline{2-4}
& Total time & Error (\%) & Total time \\
\hline
British & 58hrs & 30.33\% & 253.07s \\
\hline
Boston & 61hrs & 31.76\% & 245.26s \\
\hline
\end{tabular}
\end{small}
\vspace{-5mm}
\end{table}



\subsection{Performance in 2D}
\label{subsec:2D}
In this section, we present experimental results for 2D datasets.

\subsubsection{Effect of $\mu$}

\begin{figure}[h]
\vspace{-5mm}
\centering \subfloat[]{\includegraphics[trim = 20mm 24mm 14mm
10mm,clip,width=0.25\textwidth]{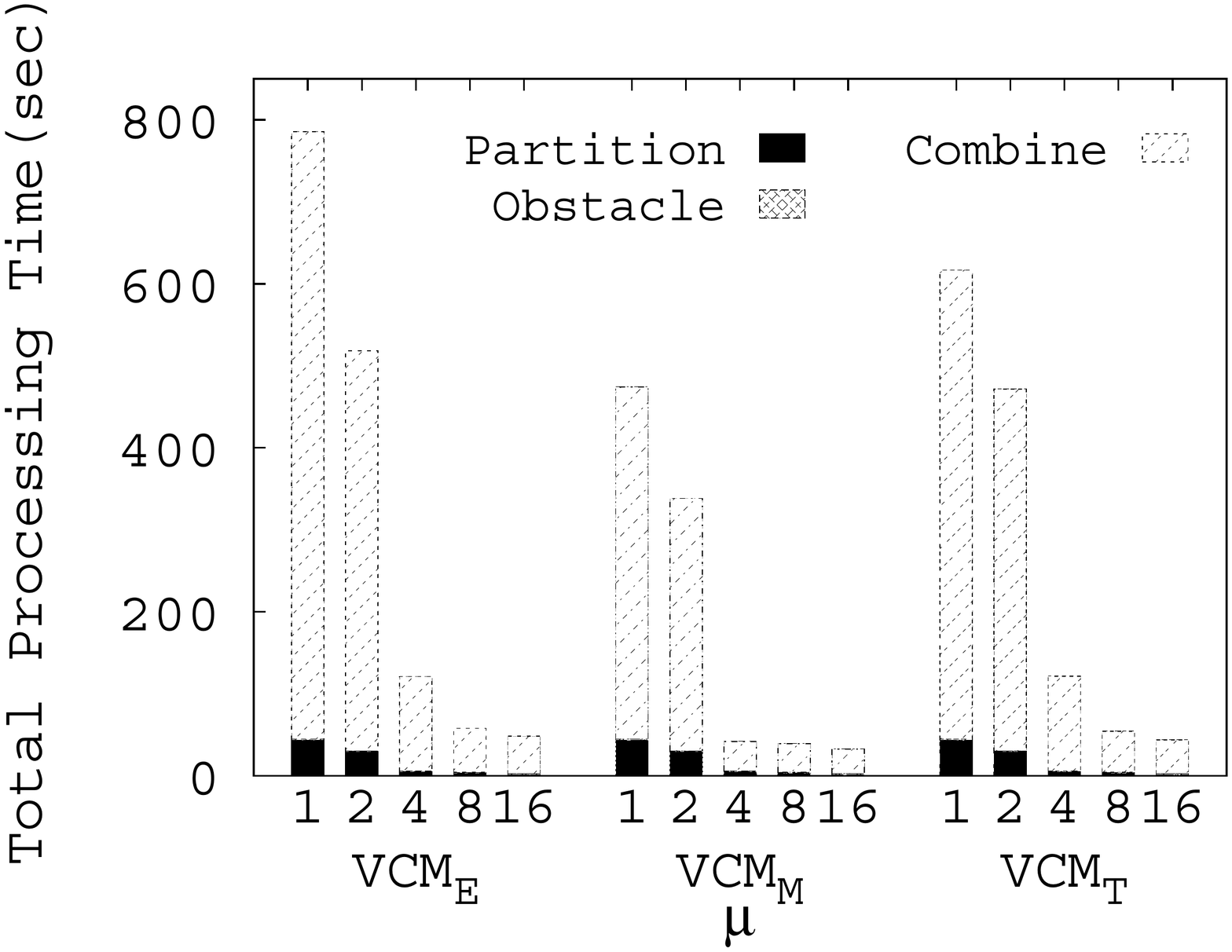}}
\subfloat[]{\includegraphics[trim = 20mm 24mm 14mm
10mm,clip,width=0.25\textwidth]{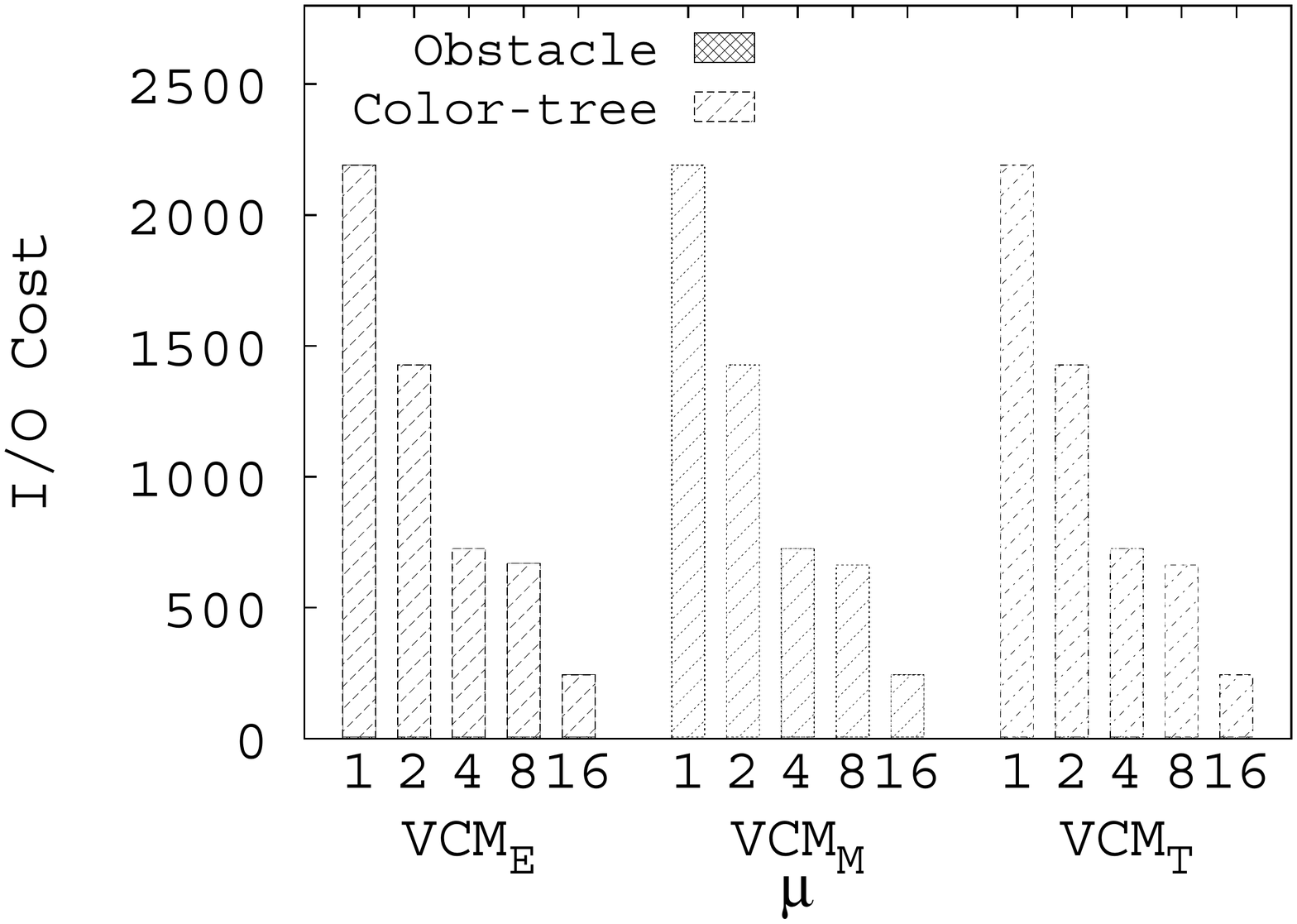}}
\vspace{-2.5mm}
\caption{Effect of $\mu$ in 2D British (a-b) dataset} \label{fig:ar_2D} \vspace{-5mm}
\end{figure}

\label{subsec:ar_variation}
\begin{table}
\centering
\vspace{-5mm}
\caption{Error in $VCM_M$ and $VCM_T$ for
varying $\mu$ in 2D}
\label{table:error2D_ar}
\begin{small}
\begin{tabular}{|l|p{1cm}|l|l|l|l|l|}
\hline Dataset & Method & \multicolumn{5}{c|}{Error (\%)} \\ \cline{3-7}
 &  & 1 & 2 & 4 & 8 & 16 \\ \hline
\multirow{2}{*}{British} & $VCM_M$ & 2.0 & 1.64 & 4.25 & 4.42 & 5.38 \\
\cline{2-7} & $VCM_T$ & 0.01 & 0.01  & 0.01 & 3.64  & 3.91 \\
\hline
\end{tabular}
\end{small}
\end{table}

In this experiment, we vary the value of $\mu$ as 1, 2, 4, 8, and
16 arcminutes and measure the total processing time and I/Os for British datasets (Fig.~\ref{fig:ar_2D}). We also present the errors resulted in the two
approximations (Table ~\ref{table:error2D_ar}).

For British dataset, on average $VCM_M$ and $VCM_T$ are 65\% times
and 17\% times faster than $VCM_E$, respectively. \eat{With the increase in $\mu$ the cell size increases and consequently the I/O cost decreases for all three methods. }In general, with the increase in $\mu$, the cell size increases and the number of total partition in the dataspace decreases. Moreover, larger $\mu$ yields fewer branching in the visibility region quad-tree. So, with the increase in $\mu$, total processing time and I/O cost decrease rapidly for both datasets. The total I/O cost is composed of (i) cost of partitioning the total dataspace to form the color-tree, (ii) cost of retrieving obstacles to form visibility region quad-tree, and (iii) cost of combining the color-tree and quad-tree to form the VCM. As these three costs hardly differ for all three approaches, they result into similar I/O cost. On the other hand, as with the increase in $\mu$, the area of partition cells ($A_i$) gets larger, the estimated error increases for both $VCM_M$ and $VCM_T$ (Equation~\ref{error_eq}). The average errors introduced
in $VCM_M$ and $VCM_T$ are 3.5\% and 1.5\%, respectively.

 \begin{figure}[h]
\centering
\subfloat[]{\includegraphics[trim = 20mm 24mm 14mm
10mm,clip,width=0.25\textwidth]{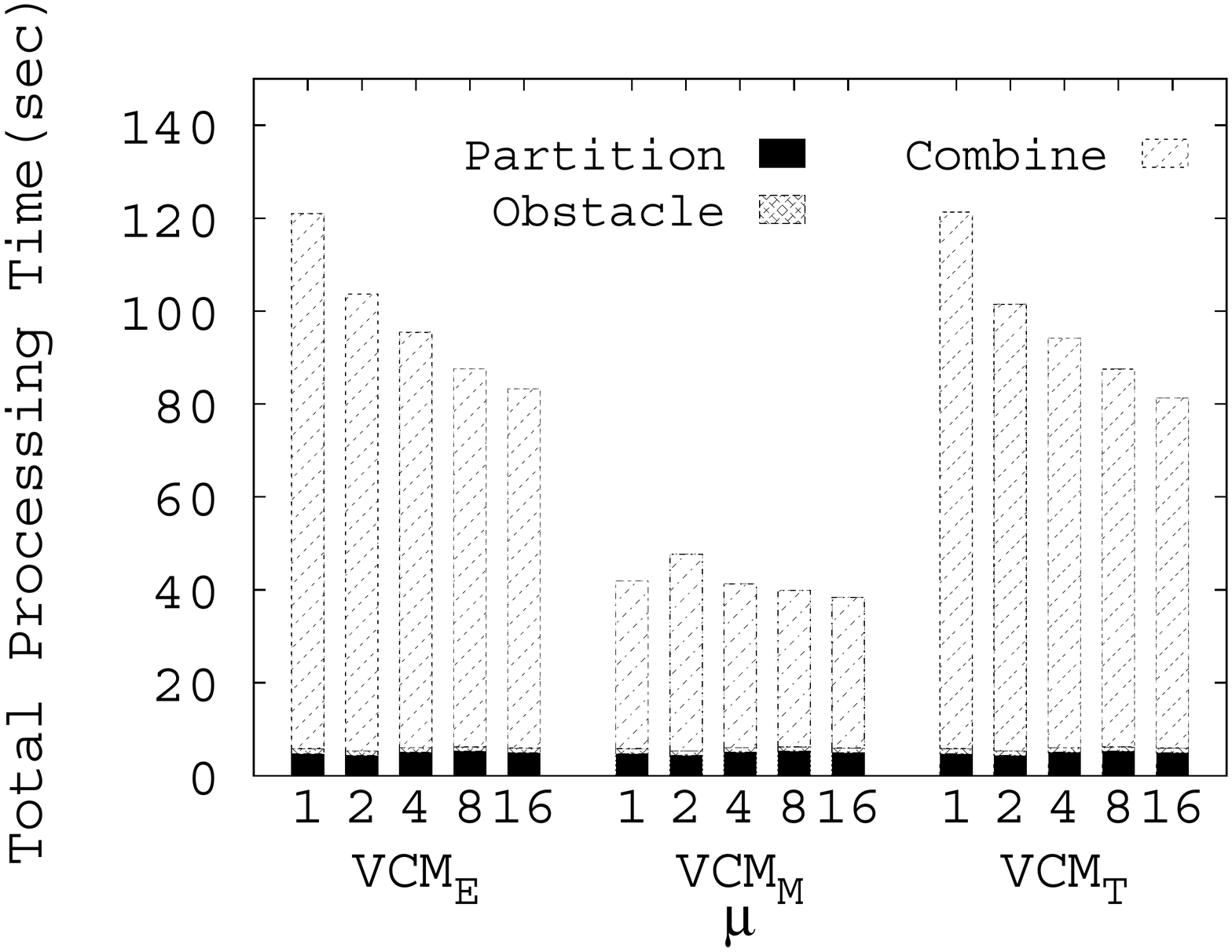}}
\subfloat[]{\includegraphics[trim = 20mm 24mm 14mm 10mm,clip,width=0.25\textwidth]{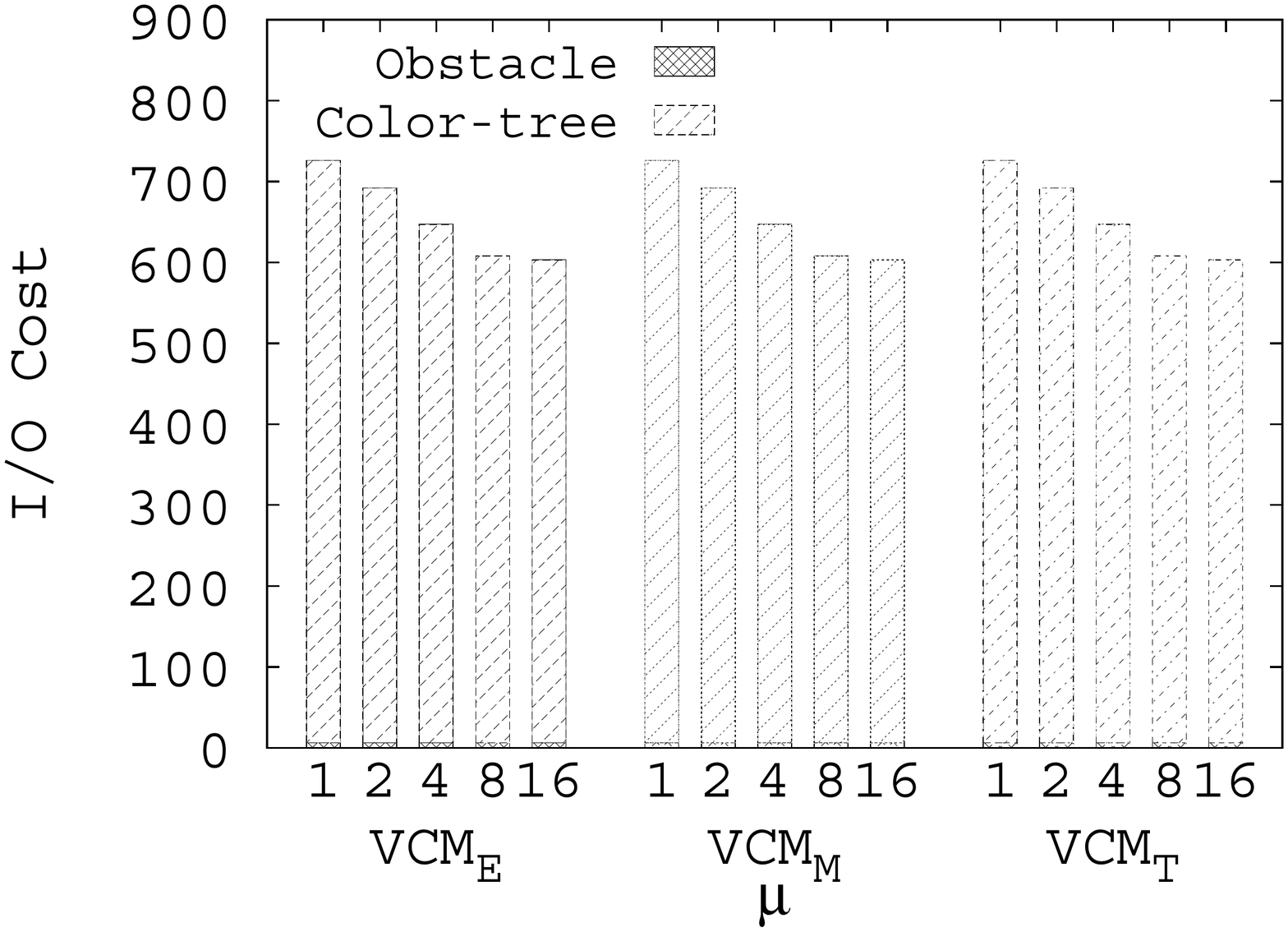}}
\vspace{-2.5mm} \caption{Effect of $\vartheta$ in 2D British (a-b)
} \label{fig:mins_variation}
\vspace{-4mm}
 \end{figure}

 \subsubsection{Effect of $\vartheta$}
\label{subsec:minsize_variation}
\begin{table}
\centering
\caption{Errors in $VCM_M$ and $VCM_T$ for $\vartheta$ in 2D}
\label{table:error2D_mb}
\begin{small}
\begin{tabular}{|l|p{1cm}|l|l|l|l|l|}
\hline Dataset & Method & \multicolumn{5}{c|}{Error (\%)} \\ \cline{3-7}
 &  & 1 & 2 & 4 & 8 & 16 \\ \hline
\multirow{2}{*}{British} & $VCM_M$ & 0.01 & 2.74 & 9.29 & 12.59 & 12.83 \\
\cline{2-7} & $VCM_T$ & 0.01 & 0.77  & 4.37 & 6.26  & 7.54 \\
\hline
\end{tabular}
\end{small}
\vspace{-14pt}
\end{table}

In this experiment, we vary the quad-tree block size ($\vartheta$)
as 1, 2, 4, 8, and 16 times of the minimum threshold of a
quad-tree block. The results are
presented in Fig.~\ref{fig:mins_variation} and Table~\ref{table:error2D_mb}. \eat{As the value of $\vartheta$ is
increased, the performance of the solution in terms of I/O cost
and total processing time increases in the cost of decreasing
accuracy. More specifically, as the minimum threshold of block
size for partition gets larger, the error rate calculated
according to Equation~\ref{error_eq} increases.}

For British dataset, $VCM_M$ is approximately 2.3 times faster than both $VCM_T$
and $VCM_E$. As explained earlier, $VCM_M$ and $VCM_T$ yield I/O costs similar to $VCM_E$. On average, $VCM_M$ and $VCM_T$ introduce 7.5\% and 3.8\% errors, respectively. These
amounts of errors do not have noticeable impact on many practical
applications. We observe from Equation~\ref{eq} that the cells
generated near the target are very small.
Therefore the approximations of the cells close to the target are
almost similar to the corresponding exact cells. The cell size
increases with the increase of distance and a significant portion of
errors is introduced for these distant cells. As the visibility
of the target from distant cells is insignificant in most of the
cases, the error in such cells is tolerable.

%
In general, with the increase in $\vartheta$, the introduced
errors in $VCM_M$ and $VCM_T$ increase as a larger quad-block size
approximates the cells with lesser accuracy. But with the increase
in $\vartheta$, the total processing time and I/O cost reduce
significantly. Hence for applications that can tolerate reduced
accuracy, a large $\vartheta$ can result into better performance.

\subsection{Performance in 3D}
\label{subsec:3D}
In this section, we present the experimental
results for 3D British and Boston datasets by varying different parameters.

\begin{table}
\centering
\vspace{-7mm}
\caption{Errors in $VCM_M$ and $VCM_T$ for $\mu$ in 3D}
\label{table:error3D_mu}

\begin{small}
\begin{tabular}{|l|p{1cm}|l|l|l|l|l|}
\hline Dataset & Method & \multicolumn{5}{c|}{Error (\%)} \\ \cline{3-7}
 &  & 1 & 2 & 4 & 8 & 16 \\ \hline
\multirow{2}{*}{British} & $VCM_M$ & 4.56 & 8.02 & 8.11 & 10.46 & 15.25 \\
\cline{2-7} & $VCM_T$ & 4.1 & 7.63 & 7.76 & 8.89  & 13.89 \\
\hline

\multirow{2}{*}{Boston} & $VCM_M$ & 3.31 & 5.03 & 8.02 & 9.12 & 10.05 \\
\cline{2-7} & $VCM_T$ & 3.12 & 4.70 & 7.62 & 8.57 & 9.79 \\
\hline
\end{tabular}
\end{small}
\end{table}

\subsubsection{Effect of $\mu$}
\label{subsec:3dar_variation}

\begin{figure}[h]
\centering
\vspace{-8mm}
\subfloat[]{\includegraphics[trim = 20mm 24mm 14mm
10mm,clip,width=0.25\textwidth]{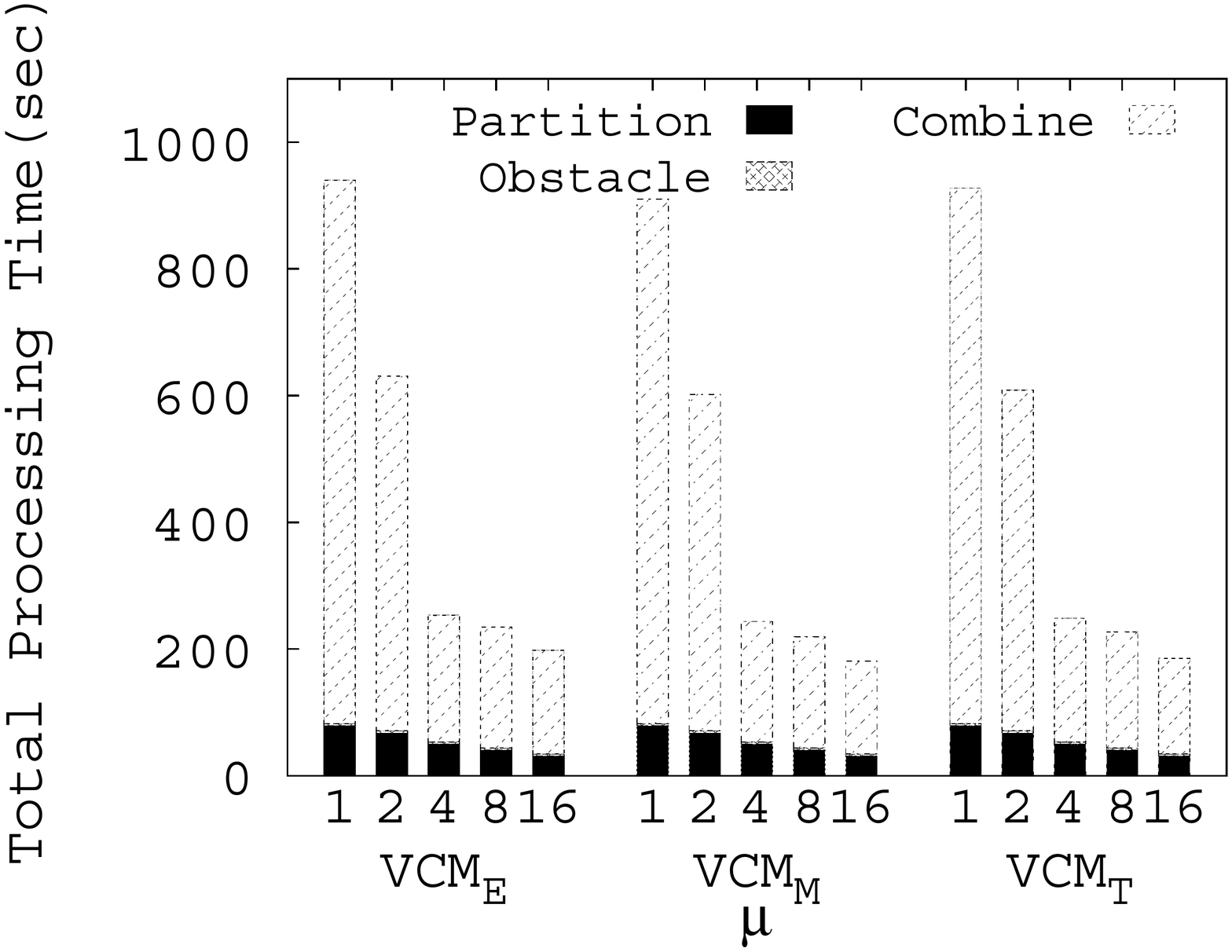}}
\subfloat[]{\includegraphics[trim = 20mm 24mm 14mm
10mm,clip,width=0.25\textwidth]{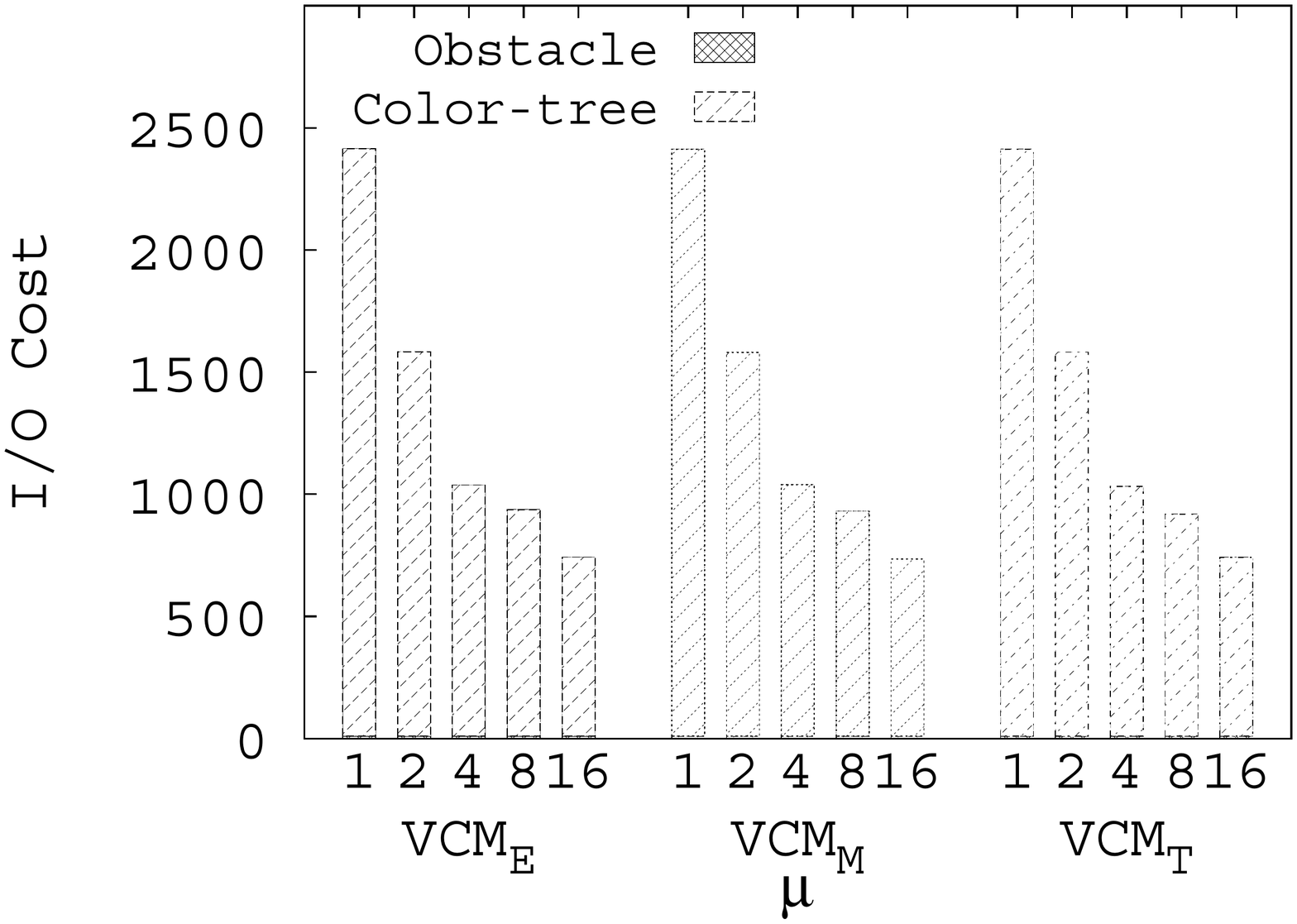}}
\vfill
\vspace{-6mm}
\subfloat[]{\includegraphics[trim = 20mm 24mm 14mm
10mm,clip,width=0.25\textwidth]{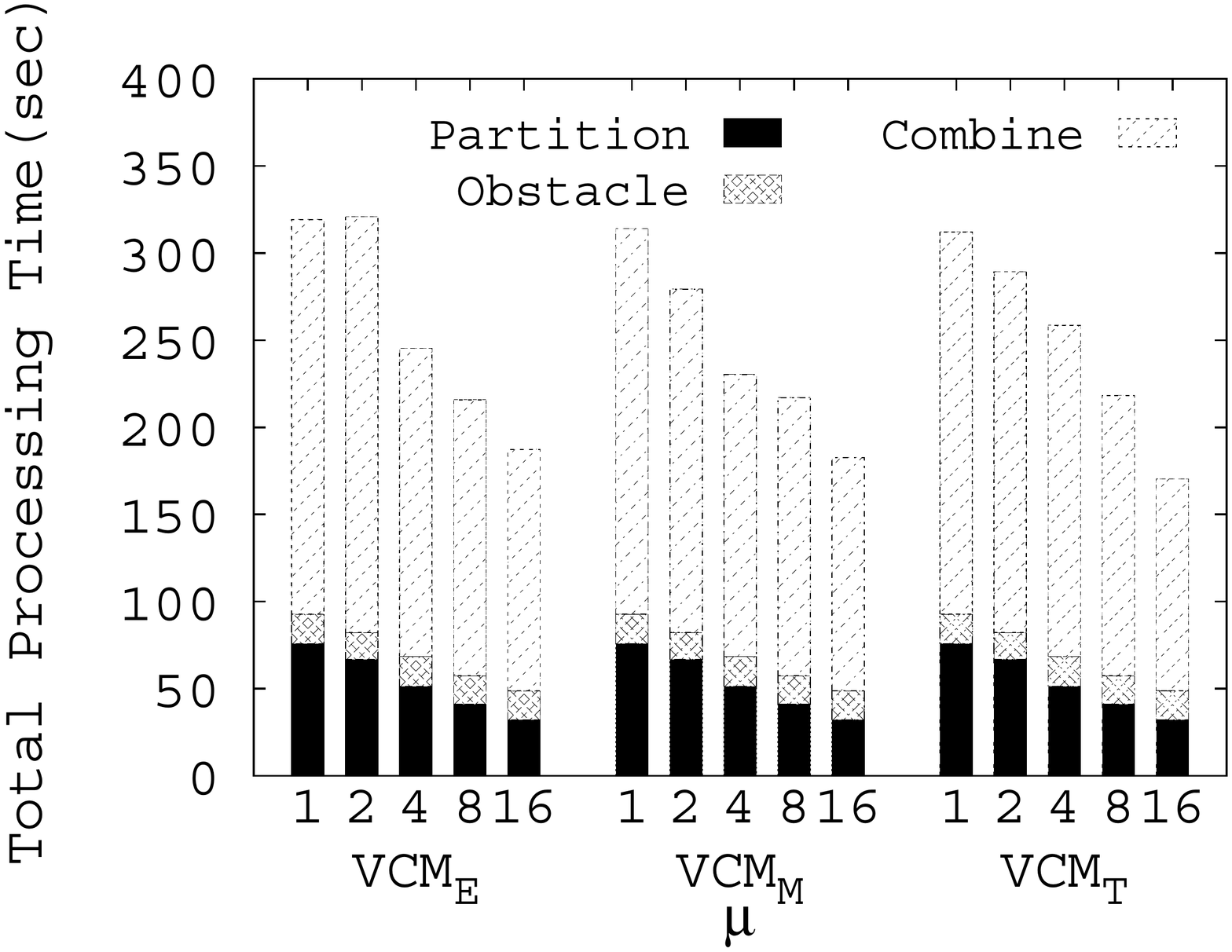}}
\subfloat[]{\includegraphics[trim = 20mm 24mm 14mm
10mm,clip,width=0.25\textwidth]{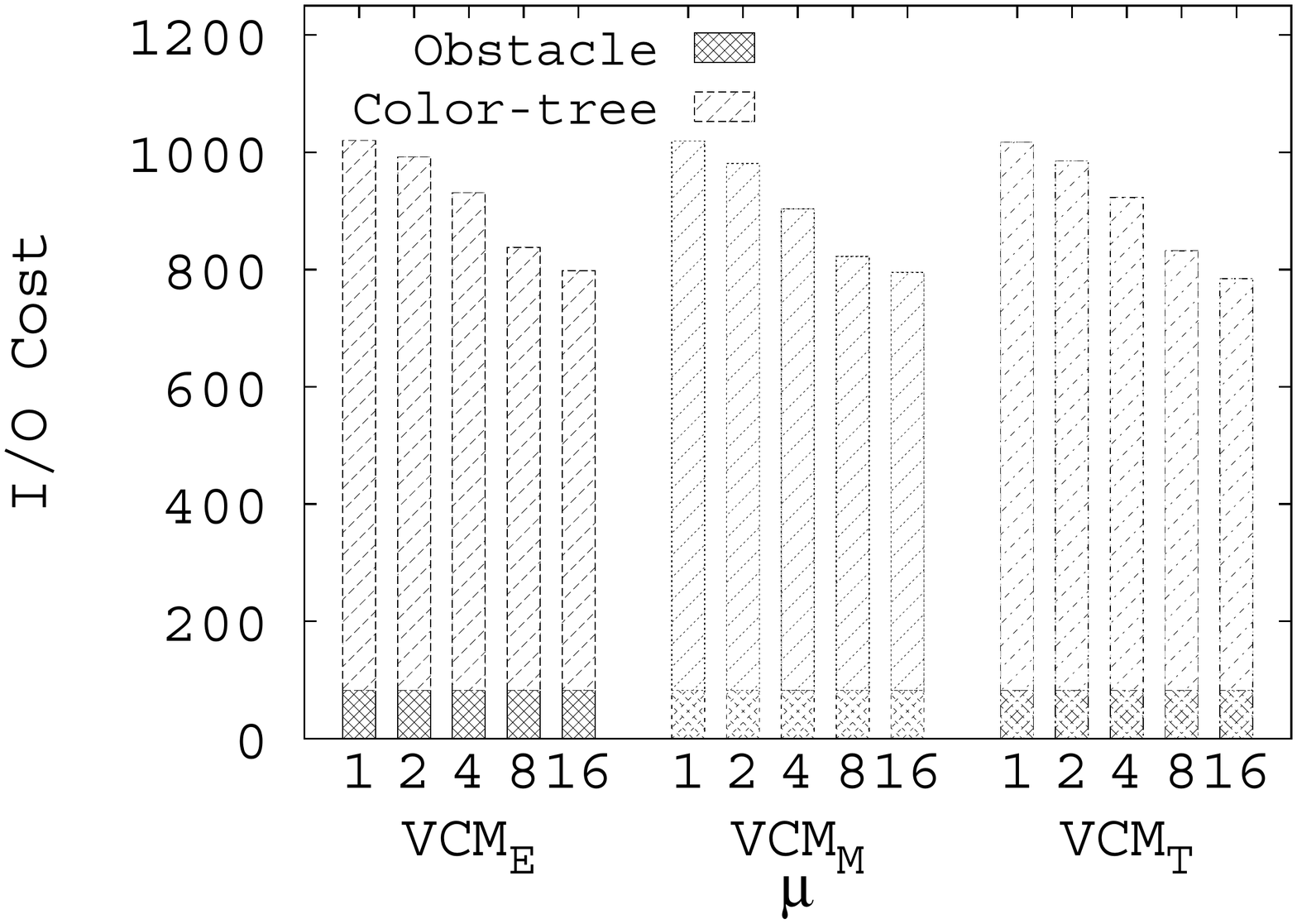}}
\vspace{-3.0mm} \caption{Effect of $\mu$ in 3D British (a-b) and
Boston (c-d)} \label{fig:ar_3D} \vspace{-4mm}
\end{figure}

In this experiment, we vary the value of $\mu$ as 1, 2, 4, 8, and
16 arcminutes and measure the total processing time and I/Os for British and Boston datasets (Fig.~\ref{fig:ar_3D}). We also present the errors resulted in the two approximation methods (Table~\ref{table:error3D_mu}).

\eat{In general, with the increase in $\mu$, the cell size increases and number of total partition in the dataspace decreases. Moreover, larger $\mu$ yields fewer branching in the visibility region quad-tree. As a result, as $\mu$ increases, it requires fewer I/O access and shorter processing time. So, with the increase in $\mu$, total processing time and I/O cost decreases rapidly for both datasets. On the other hand, as with the increase in $\mu$, the area of partition cells ($A_i$) gets larger, from Equation~\ref{error_eq} the estimated error increases for both $VCM_M$ and $VCM_T$.}

For British dataset, on average $VCM_M$ and $VCM_T$ run only 5\% and 3\%
faster than $VCM_E$, respectively. The difference in I/O cost, i.e., the number of pages accessed is also negligible among the three methods. As discussed earlier in Section~\ref{sec:approach} \eat{(i) the partitioning and (ii) visible region construction phases are identical for all three methods and }the three methods differ only in the final phase, i.e., combining the outcome of previous two steps
to produce the VCM. As we consider only full visibility of the target object from a partition cell, a huge number of cells in the dataspace fall into the obstructed region. So in the final phase the three solutions perform similarly as they have to combine the same color-tree and visible region quad-tree. The average errors introduced in $VCM_M$ and $VCM_T$ are 9\% and 8\%, respectively.

Although the results for both British and Boston datasets follow similar pattern, Boston dataset causes much smaller I/O and computational cost. Because, in case of the densely populated Boston dataset, most of the obstacles are pruned rapidly during the visible region quad-tree formation. So there are lesser number of visibility computation and page access to construct VCM in Boston dataset. The processing time of $VCM_M$ is
on average 8\% faster than that of $VCM_E$ and $VCM_T$. The average error introduced in $VCM_M$ is 7\%, whereas the more accurate $VCM_T$ yields 6\% error.

The I/O costs in the obstacle \emph{retrieval} phase are same in
all three methods. Thus in subsequent sections, we show only
the I/O cost required to \emph{combine} the quad-tree and the
color-tree (i.e., the final phase of constructing the VCM).

 \begin{figure}[h]
\centering \subfloat[]{\includegraphics[trim = 20mm 24mm 14mm
10mm,clip,width=0.25\textwidth]{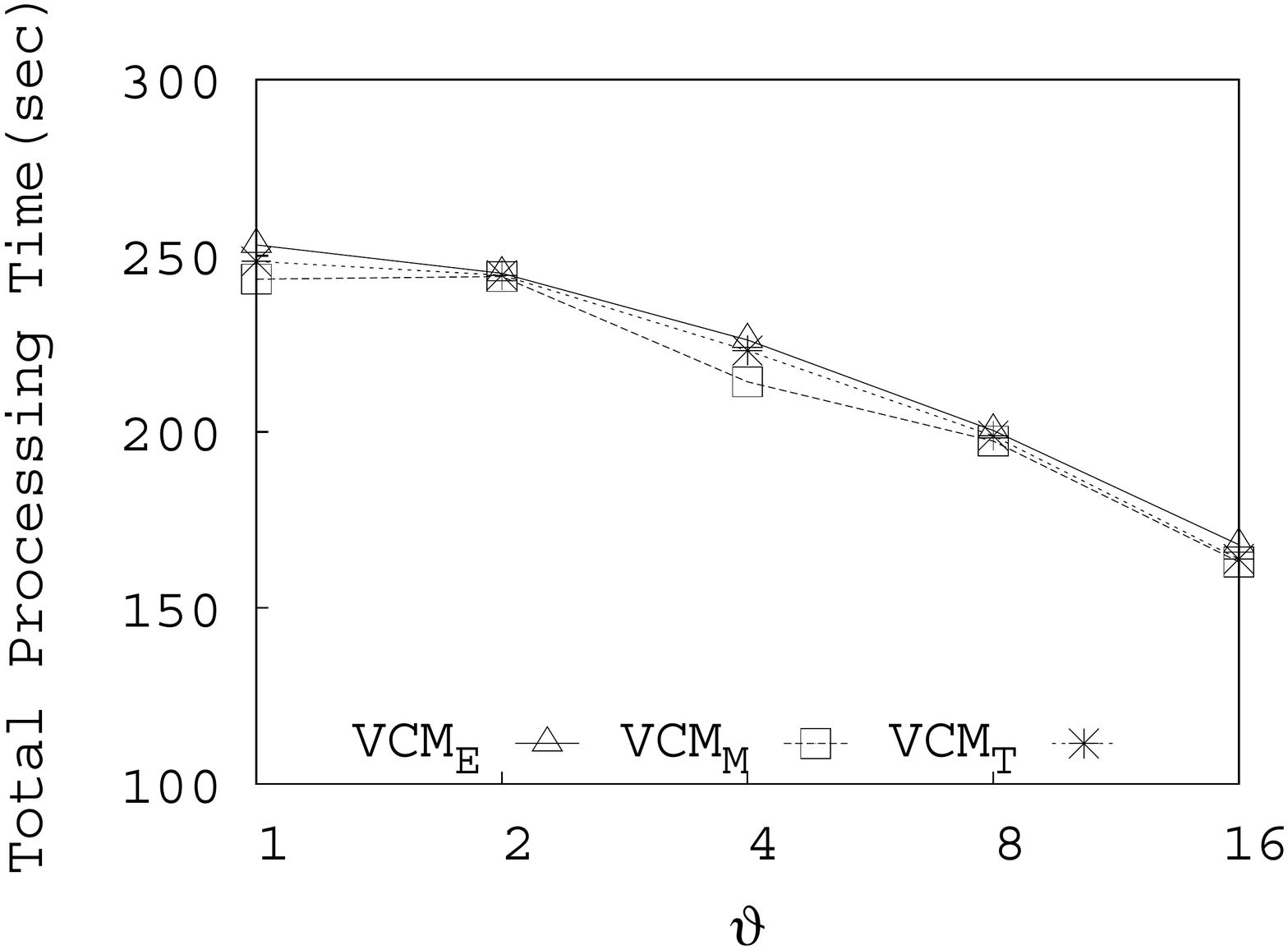}}
\subfloat[]{\includegraphics[trim = 20mm 24mm 14mm 10mm,clip,width=0.25\textwidth]{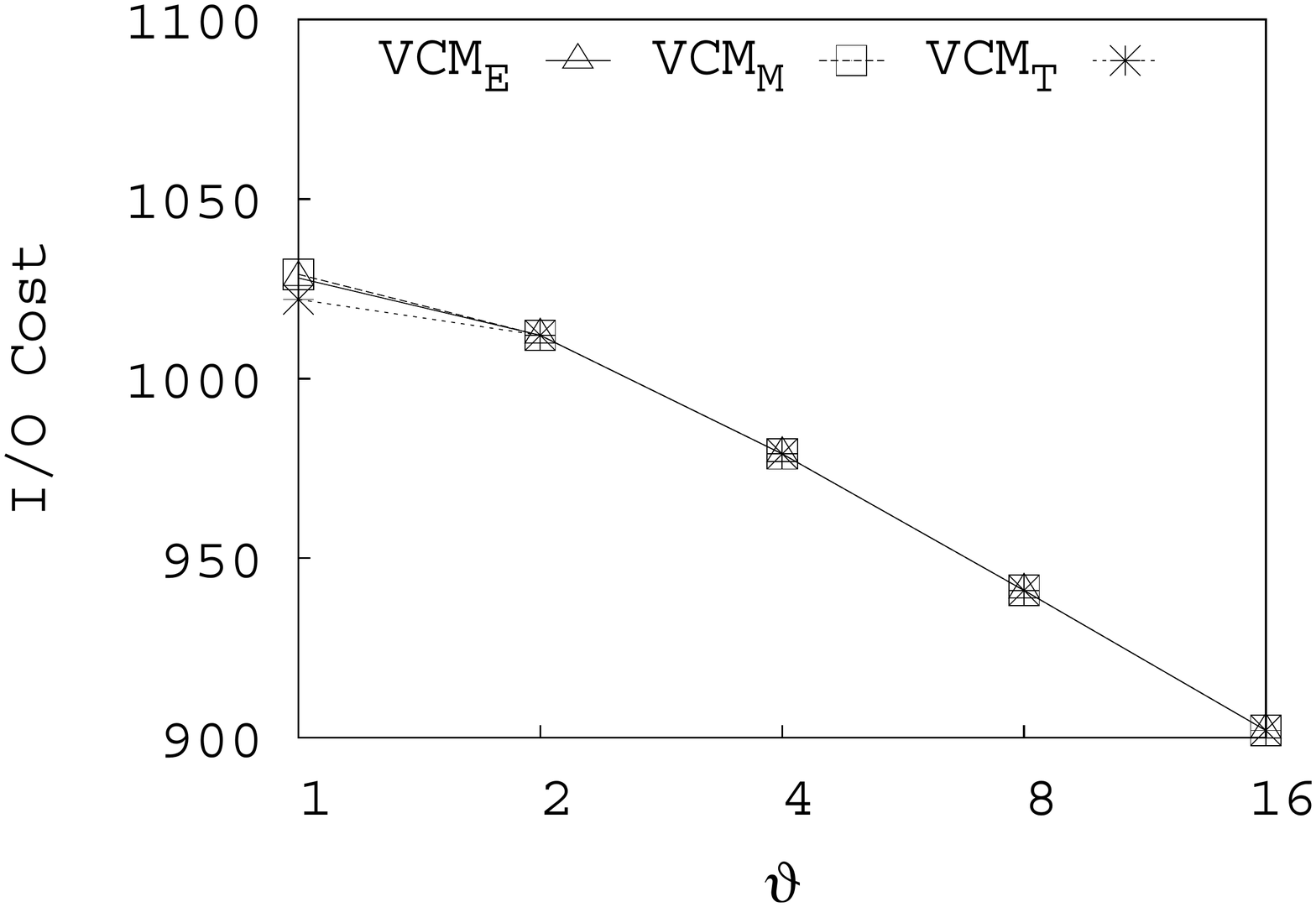}}
\vfill
\vspace{-6mm}
\subfloat[]{\includegraphics[trim = 20mm 24mm 14mm 10mm,clip,width=0.25\textwidth]{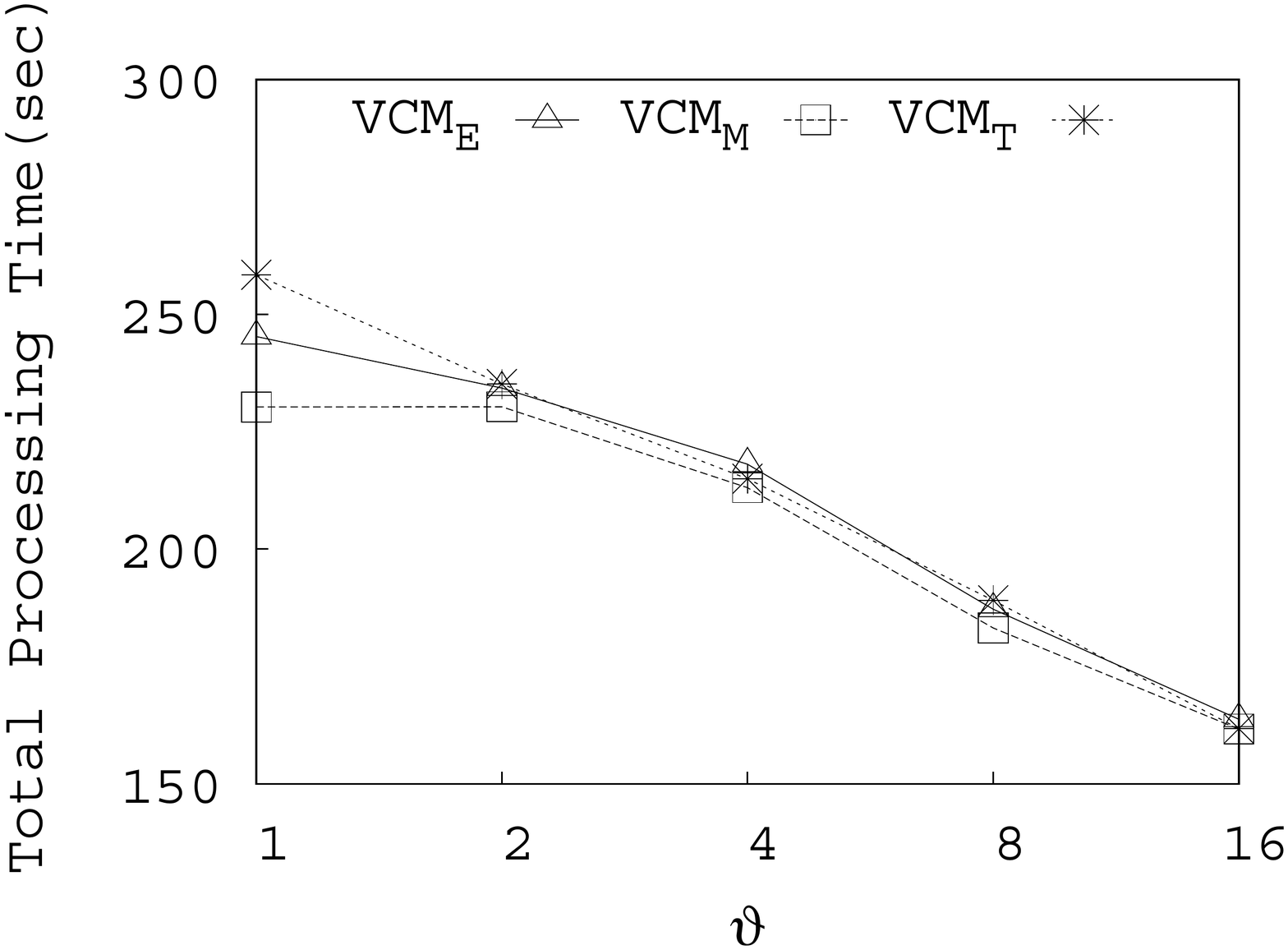}}
\subfloat[]{\includegraphics[trim = 20mm 24mm 14mm 10mm,clip,width=0.25\textwidth]{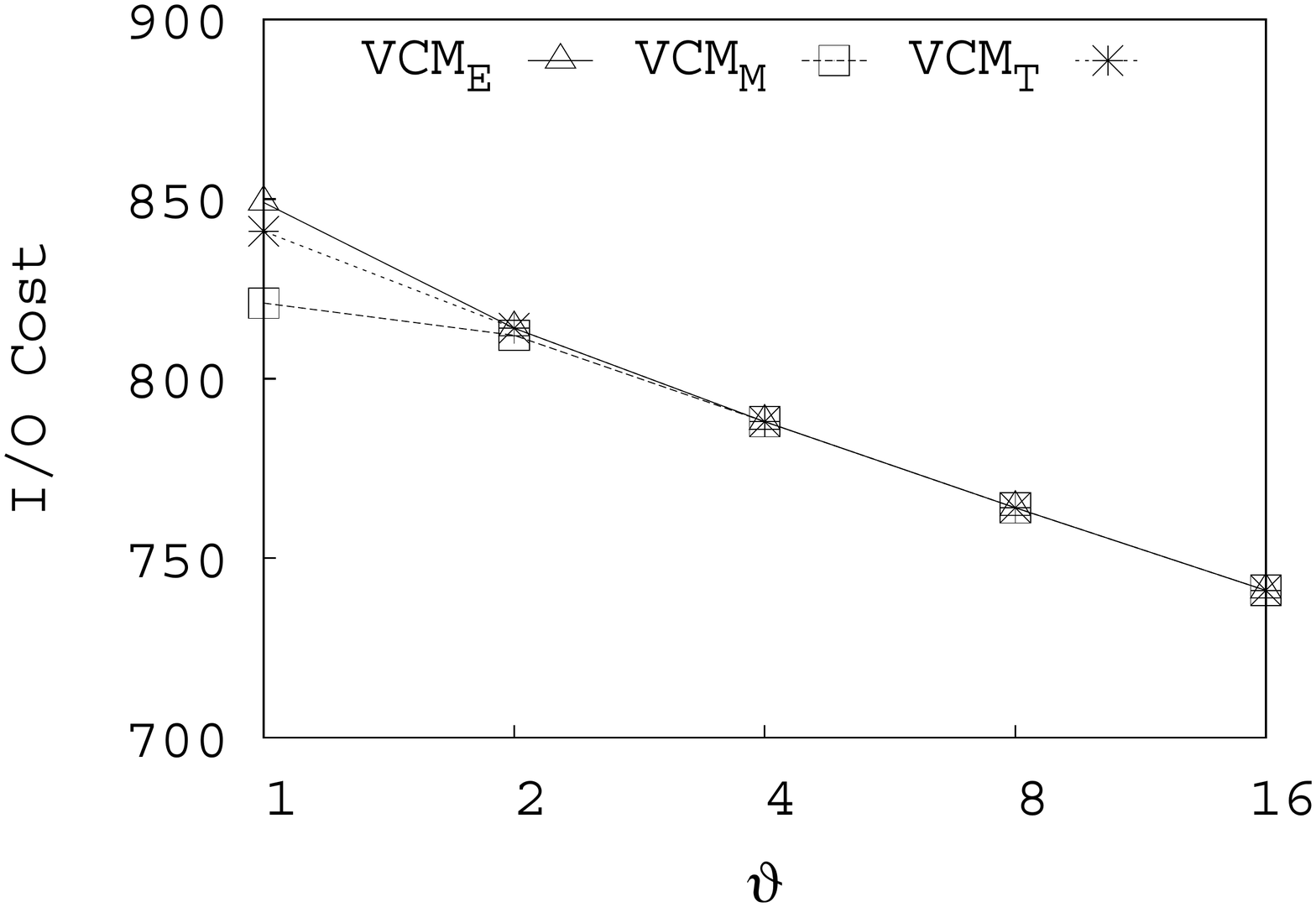}}
\vspace{-2.5mm} \caption{Effect of $\vartheta$ in 3D British (a-b)
and Boston (c-d)} \label{fig:mins_variation3d}
\vspace{-7mm}
 \end{figure}

\begin{table}
\centering
\caption{Errors in $VCM_M$ and $VCM_T$ for $\vartheta$ in 3D}
\label{table:error3D_var}
\begin{small}
\begin{tabular}{|l|p{1cm}|l|l|l|l|l|}
\hline Dataset & Method & \multicolumn{5}{c|}{Error (\%)} \\ \cline{3-7}
 &  & 1 & 2 & 4 & 8 & 16 \\ \hline
\multirow{2}{*}{British} & $VCM_M$ & 8.11 & 9.31 & 10.15 & 11.83 & 14.96 \\
\cline{2-7} & $VCM_T$ & 7.76 & 9.00 & 9.86 & 11.05 & 12.72 \\
\hline

\multirow{2}{*}{Boston} & $VCM_M$ & 8.02 & 9.08 & 9.38 & 9.82 & 11.03 \\
\cline{2-7} & $VCM_T$ & 7.62 & 8.82 & 9.06 & 8.98 & 10.29 \\
\hline
\end{tabular}
\end{small}
\vspace{-20pt}
\end{table}

 \subsubsection{Effect of $\vartheta$}
\label{subsec:3dminsize_variation}
In this experiment, we vary the quad-tree block size $\vartheta$
as 1, 2, 4, 8, and 16 times of the minimum threshold of a
quad-tree block and show the results in Fig.~\ref{fig:mins_variation3d} and Table~\ref{table:error3D_var}.

\eat{Generally, as the value of $\vartheta$ is
increased, the performance of the solution in terms of I/O cost
and total processing time increases in the cost of decreasing
accuracy. Because with the increase in quad-tree block size, the cost to merge the quad-tree with the color-tree reduces and larger quad-tree blocks results into larger error.}

For British dataset, the computational and I/O costs are similar for all three methods, e.g., on average $VCM_M$ runs only 3\% faster than $VCM_E$. Because, similar to the case of varying $\mu$, the three phases of constructing VCM differ slightly for these methods. The average error introduced in $VCM_M$ and $VCM_T$ is about 10\%. The results derived from Boston dataset are similar to that of British dataset. $VCM_M$ runs about 3\% faster than both $VCM_E$ and $VCM_T$. The average errors introduced in the approximation methods are about 9\%.

\eat{In general, with the increase in $\vartheta$, the introduced
errors in $VCM_M$ and $VCM_T$ increase as a larger quad-block size
approximates the cells with lesser accuracy. But with the increase
in $\vartheta$, the total processing time and I/O cost reduce
significantly. Hence for applications that can tolerate reduced
accuracy, a large $\vartheta$ can result into better performance.}

As the errors introduced by two approximations show similar trends
for $A_Q$, FOV, and $L_T$, we show only the total processing time
and the I/O cost in the subsequent sections.

%

\begin{figure}[h]
\centering
\vspace{-5mm}
\subfloat[]{\includegraphics[trim = 20mm 24mm 14mm
10mm,clip,width=0.25\textwidth]{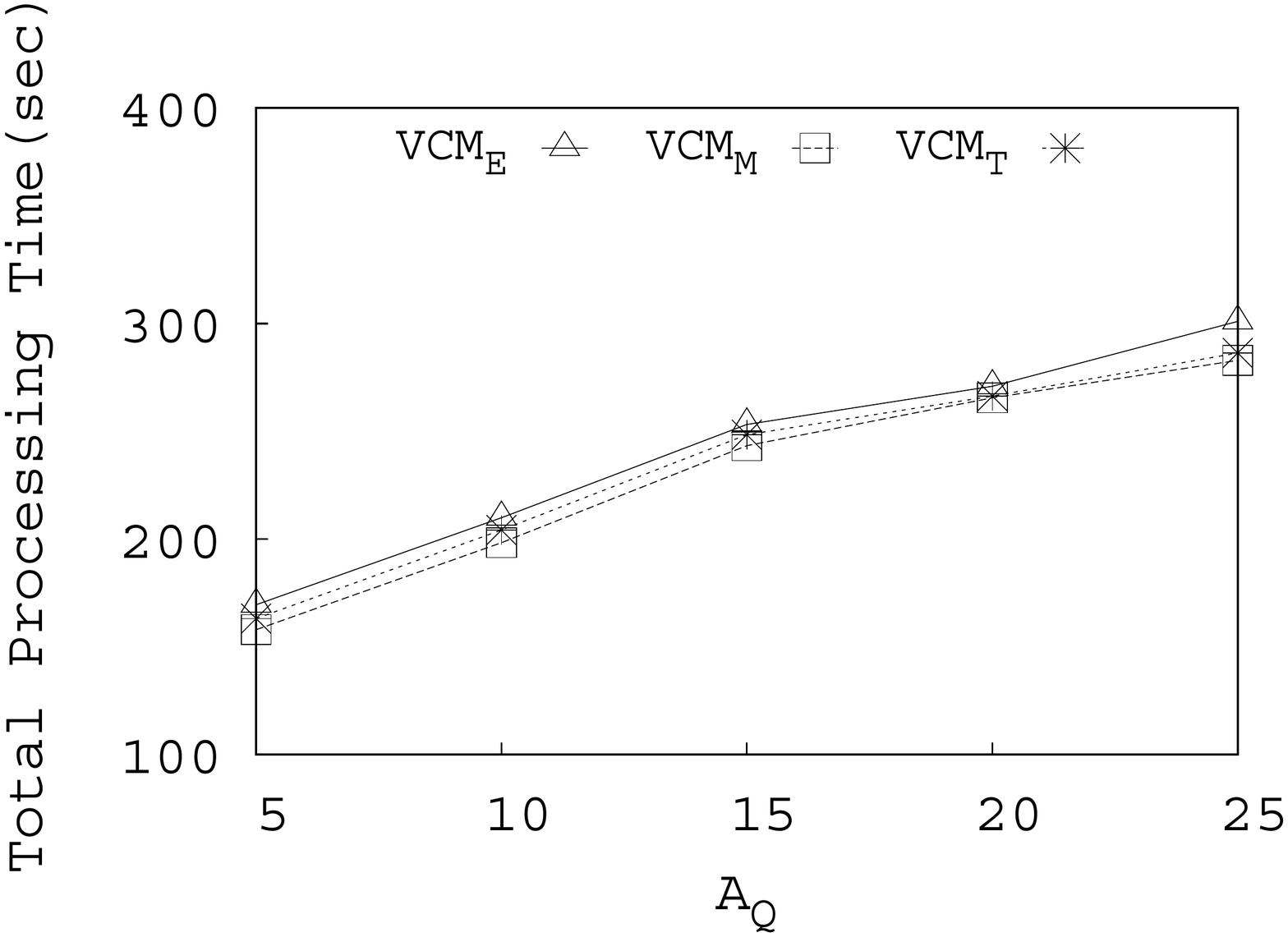}}
\subfloat[]{\includegraphics[trim = 20mm 24mm 14mm
10mm,clip,width=0.25\textwidth]{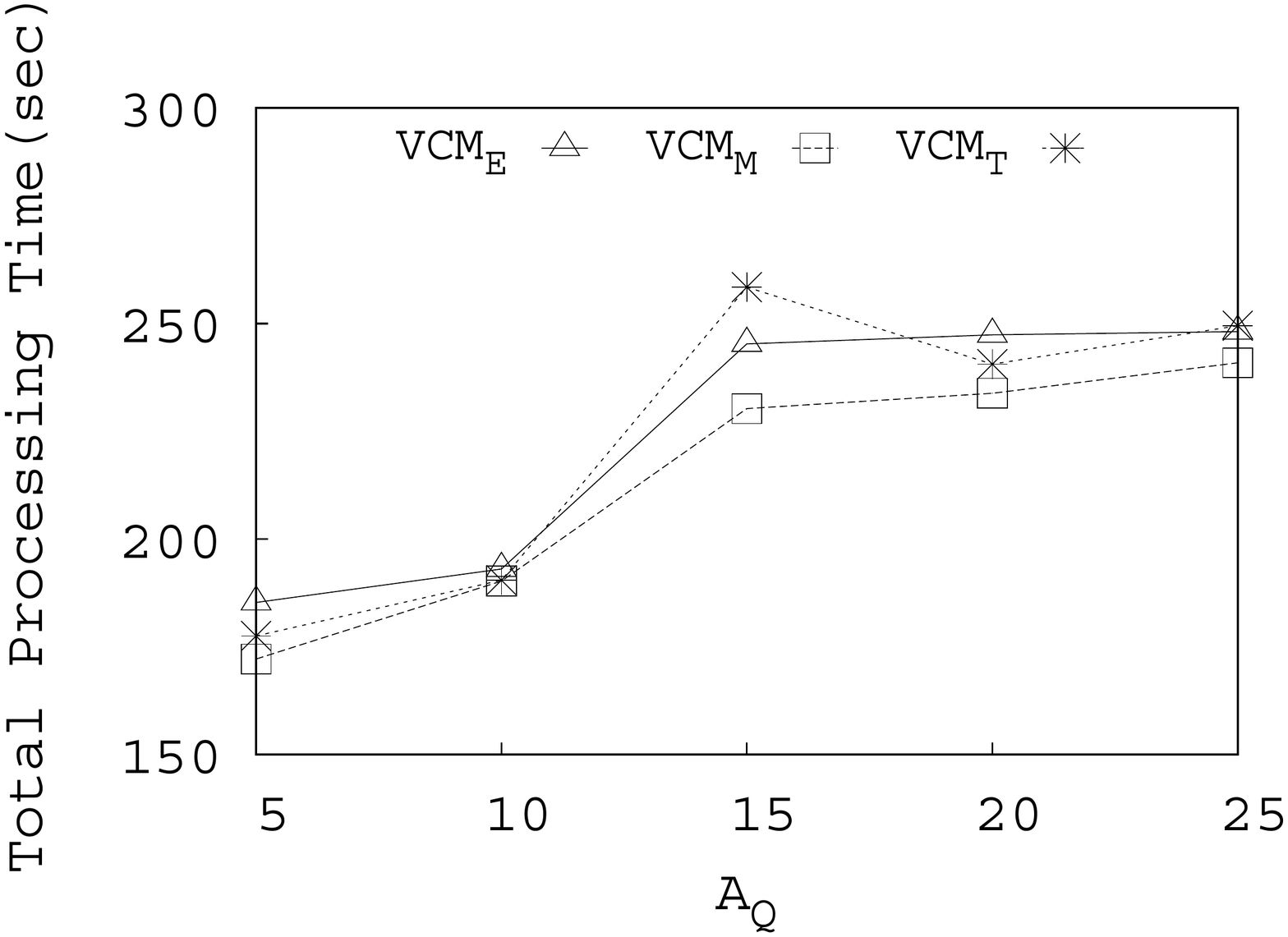}}
\vspace{-2.5mm} \caption{Effect of $A_Q$ in 3D British (a) and
Boston (b)} \label{fig:area_variation3D} \vspace{-12mm}
\end{figure}

\begin{figure}[h]
\centering \subfloat[]{\includegraphics[trim = 20mm 24mm 14mm
10mm,clip,width=0.25\textwidth]{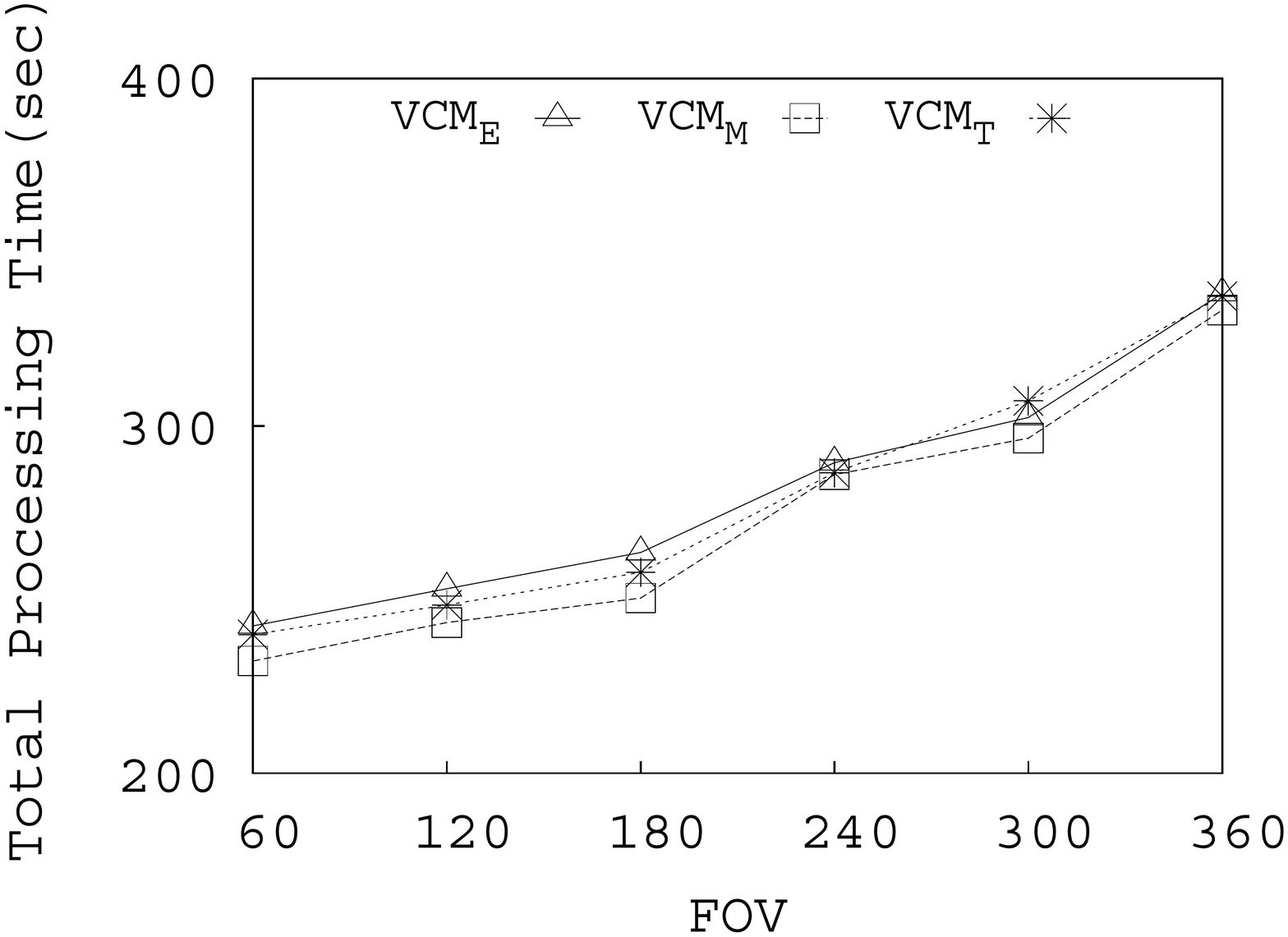}}
\subfloat[]{\includegraphics[trim = 20mm 24mm 14mm
10mm,clip,width=0.25\textwidth]{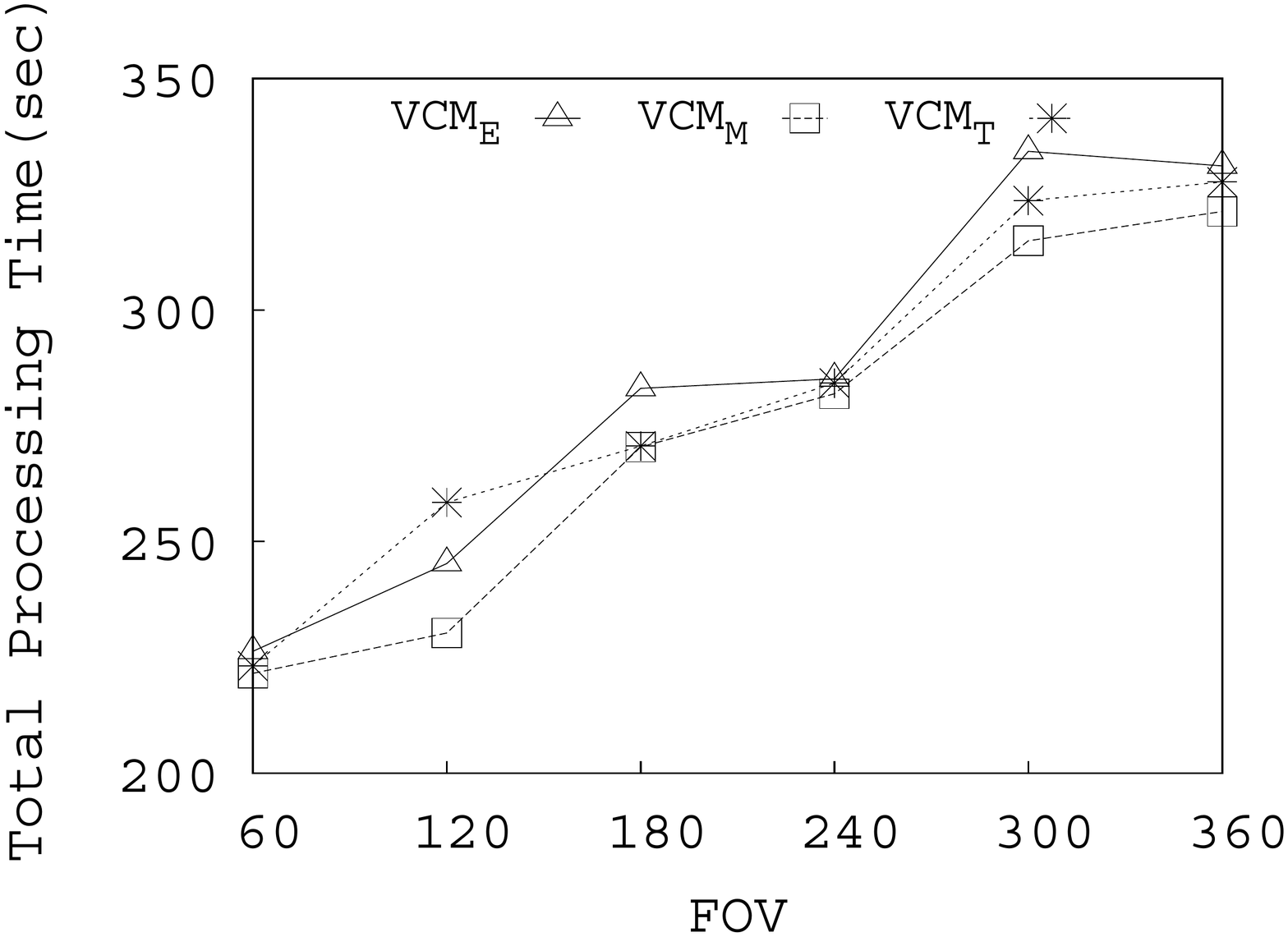}}
\vspace{-2.5mm} \caption{Effect of $FOV$ in 3D British (a) and
Boston (b)} \label{fig:fov_variation3D} \vspace{-4.5mm}
\end{figure}

\subsubsection{Effect of $A_Q$}
\label{subsec:3darea_variation}

In this experiment, we vary the query area $A_Q$ as 5\%, 10\%,
15\%, 20\%, and 25\% of the total dataspace (Fig.~\ref{fig:area_variation3D}(a)-(b)). In case of British dataset, $VCM_M$ and $VCM_T$ run 5\% and 3\% faster than $VCM_E$, respectively. For Boston dataset, $VCM_M$ runs 5\% faster than both $VCM_E$ and $VCM_T$. The I/O cost and the total processing time increase for all three methods with the increase in $A_Q$, as we need to consider higher number of cells and higher number of obstacles to construct the VCM. As the I/O cost is similar for all three methods, we show only the total processing time.

\subsubsection{Effect of FOV}
\label{subsec:3dfov_variation}

In this experiment, we vary the field of view (FOV) as 60, 120,
180, 240, 300, and 360 degrees for British
and Boston datasets (Fig.~\ref{fig:fov_variation3D}). When FOV is set to $360^o$, the
FOV covers the surrounding space in all directions. We observe
that the processing time and I/Os increase with the
increase of FOV, which is expected.

Like the previous cases, the total processing time and I/O cost are similar for all three methods. \eat{For British dataset, on average $VCM_M$ results into 10\% error while the more accurate $VCM_T$ results into 7\% error. }For British and Boston dataset, $VCM_M$ runs 3\% and 4\% faster than both $VCM_E$ and $VCM_T$, respectively. With the increase of FOV from 60 to 360 degrees, the total processing time of $VCM_E$ increases by nearly 40\% and 46\% for British and Boston dataset, respectively. On the other hand, the I/O costs increase 33\% and 31\% for British and Boston dataset, respectively.

\eat{In this case, the average error introduced by $VCM_M$ and $VCM_T$ are 13\% and 10\%, respectively.}

\subsubsection{Effect of $L_T$}
\label{subsec:3dl_variation}

%

In this experiment, we vary the length of target $L_T$ as 5\%,
10\%, 15\%, 20\%, and 25\% of the total length of the dataspace. In general, with the change in $L_T$, no significant change in performance is observed for any of the datasets (not shown). \eat{

both the processing time and
I/Os increase with the increase of target length, as the
visible area of the dataspace becomes larger for a bigger target.

In case of British dataset, $VCM_M$ is 3\% faster than both $VCM_E$ and $VCM_T$. The average error introduced by $VCM_M$ and $VCM_T$ are 9\% and 7\%, respectively. For Boston dataset, $VCM_M$ runs 4\% faster than both $VCM_E$ and $VCM_T$. On average $VCM_M$ yields 10\% error, while $VCM_T$ yields 7\% error.}

\subsubsection{Effect of varying $D_S$}
\label{subsubsec:3dobstacle variation}

 \begin{figure}[h]
\centering
\vspace{-7.4mm}
\subfloat[]{\includegraphics[trim = 20mm 24mm 14mm 10mm,clip,width=0.25\textwidth]{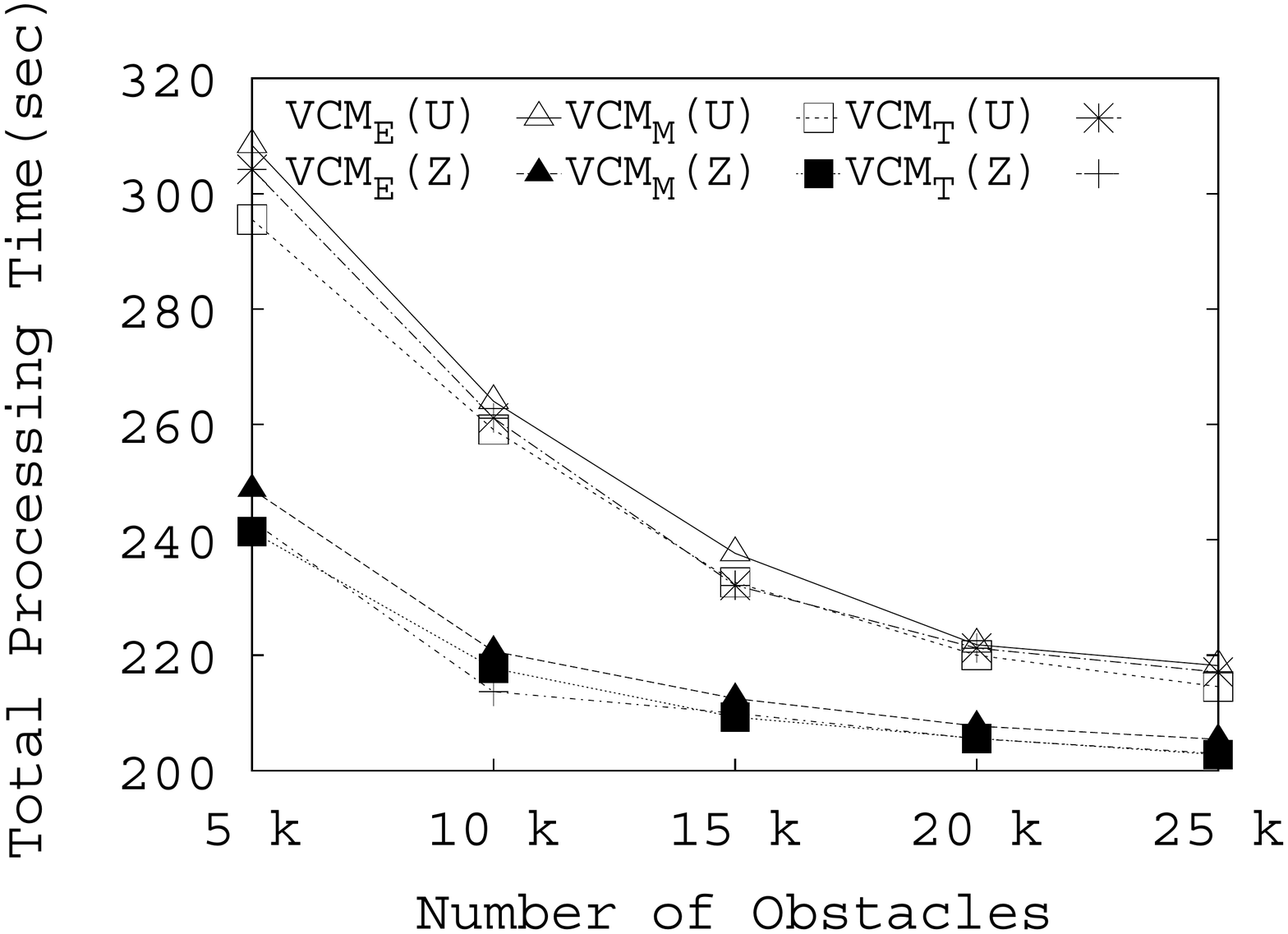}}
\subfloat[]{\includegraphics[trim = 20mm 24mm 14mm
10mm,clip,width=0.25\textwidth]{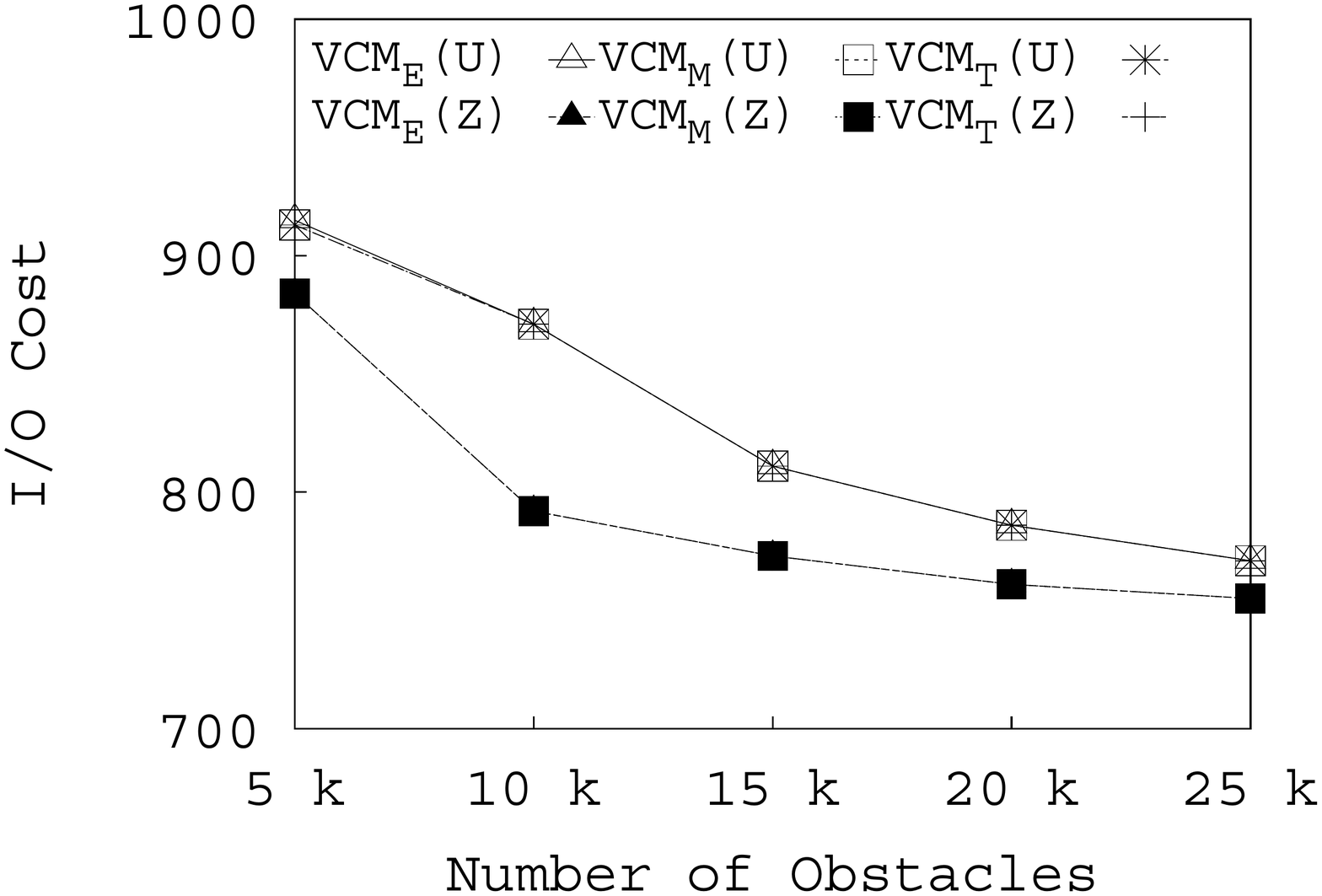}}
\vspace{-2.5mm} \caption{Effect of varying dataset size in 3D (a-b)} \label{fig:obstacle_variation3d}
\vspace{-4mm}
 \end{figure}

 In this experiment we vary the number and distribution of obstacles and measure the performance of our approximation methods in terms of I/O cost, total processing time, and approximation errors. We vary $D_S$ as a set of 5k, 10k, 15k, 20k, and 25k obstacles while keeping the other parameters at their default values. We consider both \emph{Uniform} (U) and \emph{Zipf} (Z) distributions of the obstacles. \eat{We take \emph{Uniform} (U) and \emph{Zipf} (Z) distribution of the obstacles of British dataset in 3D space.}

 In general, as the number of obstacles increases, the area of the dataspace gets more obstructed. So during the final phase of VCM construction (i.e., combining color-tree and visible region quad-tree), computational costs and I/O costs get reduced. Consequently, the overall costs decrease with the increase in $D_S$ for both Uniform and Zipf distribution. In case of Uniform and Zipf distributions, as the number of obstacles is varied from 5k to 25k, the total processing time decreases nearly 40\% and 20\%, respectively for all three methods. On the other hand, the I/O costs decrease approximately 18\% and 17\% with the increase of $D_S$ for Uniform and Zipf distributions, respectively.

\eat{We have also conducted similar experiments in 2D British and Boston datasets by varying all of $\mu$, $\vartheta$, $A_Q$, FOV, $L_T$, and $D_S$. We observe the results of 2D datasets follow similar trends to that of 3D datasets. So we are omitting the results due to the brevity of presentation.}

\section{Related Works}
\label{sec: rw} The notion of visibility is actively studied in
different contexts: computer graphics and visualization
\cite{vmapDefinition,stewart1998computing} and spatial databases
\cite{SaranaVNN,Gao:2009:CON:1559845.1559906,Gao:2009:CVN:1516360.1516378}.
Most of these techniques consider visibility as a binary notion,
i.e., a point is either visible or invisible from another point.

\subsection{\emph{Visibility in Computer Graphics and Visualization}}
In computer graphics, the visibility map refers to a planar
subdivision that encodes the visibility information, i.e., which
points are mutually \emph{visible} \cite{vmapDefinition}. Two
points are mutually visible if the straight line segment
connecting these points does not intersect with any obstacle. If a
scene is represented using a planar straight-line graph, a
horizontal (vertical) visibility map is obtained by drawing a
horizontal (vertical) straight line $l$ through each vertex $p$ of
that graph until $l$ intersects an edge $e$ of the graph or
extends to infinity. The edge $e$ is horizontally (vertically)
visible from $p$. A large body of
works~\cite{bittnerOccusntree,grasset1999accurate,keeler2007spherical,bittnerWonka2003visibility,stewart1998computing}
construct such visibility maps efficiently.

Given a collection of surfaces representing boundaries of
obstacles, Tsai et al. considers visibility problem as determining
the regions of space or the surfaces of obstacles that are visible
to an observer \cite{PDErootVisibility,PDEvisibility}. They model
visibility as a binary notion and find the light and
dark regions of a space for a point light source.\eat{ Disregarding
reflection, diffraction and interference of light, they assume
light rays are diminished upon contact with a surface of an
obstacle.}

Above methods involve the computation of visible
surfaces from a viewpoint and do not consider visibility factors
such as angle and distance to quantify the visibility of a target
object.


\subsection{\emph{Visibility in Spatial Queries}}
Visibility problems studied in spatial databases usually involve
finding the nearest object to a given query point in the presence
of obstacles. In recent years, several variants of the nearest
neighbor (NN) queries have been proposed that include Visible NN
(VNN) query \cite{SaranaVNN}, Continuous Obstructed NN (CONN)
query \cite{Gao:2009:CON:1559845.1559906}, and Continuous Visible
NN query \cite{Gao:2009:CVN:1516360.1516378}.


Nutanong et al. introduce an approach to find the NN
that is visible to a query point \cite{SaranaVNN}. A CNN query
finds the $k$ NNs for a moving query
point~\cite{Tao:2002:CNN:1287369.1287395}.\eat{  A CNN query
retrieves the NN of every point on a line segment returns a set of
(point, interval) tuples, such that the point is the NN of all
points in the corresponding interval.} Gao et al. propose a
variation of CNN; namely, a CONN query
\cite{Gao:2009:CON:1559845.1559906}. Given a dataset $P$, an
obstacle set $O$, and a query line segment $q$ in a 2D space, a
CONN query retrieves the NN of each point on $q$
according to the obstructed distance.

The aforementioned spatial queries find the nearest object in an
obstructed space from a given query point where query results are
ranked according to visible distances from that query point.
Instead of quantifying visibility as a non-increasing function
from a target to a viewpoint, they label a particular point or
region as either visible or invisible. But for constructing a VCM
of the entire space, such binary notion is not applicable.

Recently the concept of maximum visibility query is tossed in
\cite{MOV} that considers the effect of obstacles during
quantifying visibility of a target object. They measure the
visibility of a target from a \emph{given set of query points} and
rank these query points based on the visibility measured as the
visible surface area of the target.

\eat{This approach is not applicable for constructing the VCM as
we need to compute the visibility for every viewpoint of the
dataspace by considering both the distance and the angle between a
target and a viewpoint.}


\eat{On the other hand we consider properties of lens and effects
of angle and distance between target and query points to quantify
visibility. Our proposed solution  while for the  construction of
VCM visibility should be computed for each point in the entire
dataspace. Hence their approach is not feasible for an efficient
construction of VCM. }



\section{Conclusion}
In this paper, we have proposed a technique to compute a
visibility color map (VCM) that forms the basis of many real-life
visibility queries in 2D and 3D spaces. A VCM quantifies the
visibility of (from) a target object from (of) each viewpoint of
the surrounding space and assigns colors accordingly in the
presence of obstacles. Our approach exploits the limitation of a human eye or a lens to partition the space into cells in the presence of obstacles such that the target appears same from all viewpoints inside a cell.

\eat{In this paper, we have proposed a technique to compute a
visibility color map (VCM) that forms the basis of many real-life
visibility queries in 2D and 3D spaces. A VCM quantifies the
visibility of (from) a target object from (of) each viewpoint of
the surrounding space and assigns colors accordingly in the
presence of obstacles. The key insight of our solution is considering spatially close set of viewpoints collectively
instead of treating each individual viewpoint independently that exploits the limitation of a human eye or a lens. Specifically, our solution partitions the space into cells based
on the distance and angle between the target and the viewpoint and incorporates the effects of obstacles
such that the target appears same from all viewpoints inside a cell.}

Our proposed solution significantly
outperforms the \emph{baseline} approach by at least $800$ times and \emph{six}
orders of magnitude in terms of computational time and I/O cost, respectively for both datasets.
Our conducted
experiments on real 2D and 3D datasets demonstrate the efficiency
of our approximations, $VCM_M$ and $VCM_T$. On average, the approximations $VCM_M$ and
$VCM_T$ improve the processing time by 65\% and 17\% over $VCM_E$, respectively and require almost similar I/O costs.
Both of $VCM_M$ and $VCM_T$ improve the efficiency of
$VCM_E$ by introducing only 9\% and 5\% error on average.



\begin{small}
\bibliographystyle{IEEEtran}
\bibliography{sig}  
\end{small}
\balance

\end{document}